\documentclass[aps,pre,twoside,superscriptaddress,floatfix,twocolumn,a4paper,showpacs,footinbib]{revtex4-1}
\usepackage{amsmath}
\usepackage{amsfonts}
\usepackage{amssymb}
\usepackage{graphicx,epsfig}
\usepackage{placeins}
\usepackage{pbox}

\begin{document}

\title{Scaling Green-Kubo relation and application to three aging systems}

\author{A. Dechant}
\affiliation{Dahlem Center for Complex Quantum Physics, FU Berlin, 14195 Berlin, Germany}
\author{E. Lutz}
\affiliation{Dahlem Center for Complex Quantum Physics, FU Berlin, 14195 Berlin, Germany}
\affiliation{Institute for Theoretical Physics II, University of Erlangen-N{\"u}rnberg, 91085 Erlangen, Germany}
\author{D.A. Kessler}
\affiliation{Department of Physics, Institute for Nanotechnology and Advanced Materials, Bar Ilan University, Ramat-Gan 52900, Israel}
\author{E. Barkai}
\affiliation{Department of Physics, Institute for Nanotechnology and Advanced Materials, Bar Ilan University, Ramat-Gan 52900, Israel}

\begin{abstract}
The Green-Kubo formula relates the spatial diffusion coefficient to the stationary velocity autocorrelation function. 
We derive a generalization of the Green-Kubo formula valid for systems with long-range or nonstationary correlations for which the standard approach is no longer valid. 
For the systems under consideration, the velocity autocorrelation function $\langle v(t+\tau) v(t) \rangle$ asymptotically exhibits a certain scaling behavior and the diffusion is anomalous $\langle x^2(t) \rangle \simeq 2 D_\nu t^{\nu}$.
We show how both the anomalous diffusion coefficient $D_\nu$ and exponent $\nu$ can be extracted from this scaling form.
Our scaling Green-Kubo relation thus extends an important relation between transport properties and correlation functions to generic systems with scale invariant dynamics. 
This includes stationary systems with slowly decaying power law correlations as well as aging systems, whose properties depend on the the age of the system. 
Even for systems that are stationary in the long time limit, we find that the long time diffusive behavior can strongly depend on the initial preparation of the system. 
In these cases, the diffusivity $D_{\nu}$ is not unique and we determine its values for a stationary respectively nonstationary initial state.
We discuss three applications of the scaling Green-Kubo relation: Free diffusion with nonlinear friction corresponding to cold atoms diffusing in optical lattices, the fractional Langevin equation with external noise recently suggested to model active transport in cells and the L{\'e}vy walk with numerous applications, in particular blinking quantum dots.
These examples underline the wide applicability of our approach, which is able to treat very different mechanisms of anomalous diffusion.
\end{abstract}

\maketitle

\section{Introduction \label{SEC1}}

For many central results of statistical mechanics, the stationarity of the described quantities is crucial.
Stationarity is conveniently defined in terms of the autocorrelation function $\langle A(t+\tau) A(t) \rangle$ of an observable $A$, where $\langle \ldots \rangle$ denotes an ensemble average. The quantity $A$ is called stationary if this correlation function is independent of the time $t$, called the age of the system, and only depends on the time lag $\tau$. Conversely, $A$ is called aging if the correlation function depends explicitly on the age $t$.
Generally, any system may show aging for short enough times, but for a wide class of systems, relaxation into the stationary state is exponentially fast. They can thus be fully described by their stationary behavior on time scales longer than the characteristic relaxation time.
However, this is not true for all physical systems.
Some systems may posses a stationary state but the relaxation towards this state may be slow and thus important on experimentally relevant time scales \cite{kos98,lil03,min05,sil09}.
Other systems may even exhibit aging on all experimentally accessible time scales \cite{bou92,cug94,mar04}.
The extension and generalization of known results for the stationary case to the aging one is important to better understand and characterize the behavior of aging systems \cite{bur10}.
In this work, we discuss such a generalization of the Green-Kubo formula.

The Green-Kubo formula, a central result of nonequilibrium statistical mechanics, expresses the spatial diffusion coefficient, $\langle x^2(t) \rangle \simeq 2 D_1 t$, as a time integral of the stationary velocity correlation function \cite{gre54,kub57},
\begin{equation}
D_1 = \int_0^{\infty}  {\rm d}\tau\, \langle v(t+ \tau) v(t) \rangle, \label{eq01}
\end{equation}
For a Brownian particle of mass $m$ with Stokes' friction, the velocity correlation function is exponential, $\langle v(t+\tau) v(t)\rangle = k_B T/m \exp(-\gamma^* \tau)$, and we get the Einstein relation $D_1 = k_B T/(m \gamma^*)$ linking the diffusivity and the friction coefficient $\gamma^*$ via the temperature $T$. 
The Green-Kubo formula offers a simple way of relating the diffusive properties of a system and its velocity dynamics. 
However, it is only valid as long as there exists a stationary velocity correlation function and its integral is finite. 
If, by contrast the velocity correlation function decays slowly or is nonstationary, Eq.~\eqref{eq01} needs to be generalized. 

We present a generalization of Eq.~\eqref{eq01} to superdiffusive systems, $\langle x^2(t) \rangle \simeq 2 D_{\nu} t^{\nu}$ with $\nu > 1$. 
This generalized Green-Kubo relation allows us to determine the diffusive behavior of the system by studying the scaling properties of the velocity autocorrelation. 
The scaling exponent will be seen to be related to the diffusion exponent $\nu$, while the anomalous diffusion coefficient $D_{\nu}$ is, in the spirit of the original Green-Kubo formula, expressed as the integral over a scaling function. 
Our scaling Green-Kubo relation is applicable to a wide range of systems exhibiting anomalous scale invariant dynamics. These include systems whose velocity autocorrelations are stationary but exhibit a slow power law decay, systems where the correlation function shows usual aging, i.e. the correlation time increases linearly with the age of the system, and finally systems exhibiting what we refer to as superaging, where in addition to aging, the variance of the velocity increases with time.

One important finding is that even if a system's velocity correlation function is stationary in the long time limit, the long time diffusive dynamics still may not be described by this stationary correlation function. 
This implies that aging effects can be important even in systems that are known to posses a stationary state, in particular, the diffusivity can depend strongly on the initial state of the system. 
This is in contrast to the established notion of transport coefficients as unique quantities that are independent of the initial preparation of the system. 
Such a persistence of the initial condition was previously observed by Zumofen and Klafter for a certain class of dynamical systems \cite{zum93}. 
Our scaling Green-Kubo relation is able to treat both these persistence effects in stationary systems and actual aging systems, offering new insights on the interrelation between diffusive behavior and correlations.

In the course of this work, we discuss three examples of such systems with widespread applicability. The first example is diffusion under the influence of a nonlinear friction force that is inversely proportional to the velocity. This unusual kind of friction force occurs as an effective cooling force in the semiclassical treatment of cold atoms in dissipative optical lattices \cite{cas90}. The interaction of the atoms with the lattice depends strongly on their velocity and the Sisyphus cooling mechanism described by the friction force is not effective for very fast atoms. This nonlinear behavior of the friction force on the velocity leads to anomalous dynamics including velocity correlations that decay slowly in time \cite{mar96} and exhibit superaging \cite{dec11}. Our scaling Green-Kubo relation allows to determine the spatial diffusion exponent and coefficient from the structure of the velocity correlation function. In a more general context, our results also apply to diffusion in a logarithmic potential \cite{dec11}, which has applications in a wide array of physical systems, like vortex dynamics \cite{bra00}, long range interacting systems \cite{bou05}, particles near a long charged polymer \cite{man69}, nanoparticles in a trap \cite{coh05}, self-gravitating particles \cite{cha10} or the dynamics of denaturing DNA \cite{fog07}. It has been noted by Hirschberg et al. \cite{hir11,hir12} that the distribution function evolves differently for different classes of initial conditions. By means of our scaling Green-Kubo relation, we study under which conditions this sensitivity on the initial condition carries over to the diffusion coefficient.

As our second example, we consider the fractional Langevin equation with external noise. In this case, the anomalous dynamics are due to a power law friction kernel, which describes a long range hydrodynamic memory, and power law correlated noise \cite{lut01,pot03}. The equilibrium fractional Langevin equation has been shown to provide a good description for anomalous subdiffusion in a crowded environment, like the cytoplasm of biological cells, in thermal equilibrium \cite{wei04,wei13}. In Refs.~\cite{bru09,dec13} an external nonequilibrium noise term has been introduced to describe the superdiffusive behavior due to active forces in living cells. In this superdiffusive regime, we use our scaling Green-Kubo relation to determine the diffusive properties from the velocity autocorrelation. In particular, we find that the diffusion coefficient is sensitive to whether the system is initially in the stationary state or not, indicating that the naive stationary description of the system may not provide correct results for the measured transport coefficients.

Our final example is a paradigm model for aging, the L{\'e}vy walk \cite{bou90,kla90}. In its simplest form, this model describes a particle whose velocity can have values $\pm v_0$ with random switching between them. The time between two consecutive changes in velocity is randomly selected according to a heavy tailed power law waiting time distribution, $\psi(t_{\rm w}) \sim t_{\rm w}^{-\mu-1}$ with $0 < \mu < 2$. In particular, when $0 < \mu < 1$, the mean waiting time is infinite and the velocity correlation function ages \cite{god01,bar03}. The L{\'e}vy walk has been successfully applied to various systems, including turbulence \cite{shl87}, search patterns of animals \cite{ram04} and blinking quantum dots \cite{jun01}. For the latter case, we show how our scaling Green-Kubo relation offers a straightforward way to connect the anomalous photon statistics to the intensity autocorrelations.

\section{Green-Kubo relation \label{SEC2}}

We begin by generalizing the Green-Kubo formalism to treat nonstationary power-law correlations. Our starting point is the velocity correlation function, $C_{v}(t_2 , t_1) = \langle v(t_2) v(t_1) \rangle $, which  is defined as,
\begin{equation}
C_{v}(t_2, t_1) 
= \!\int_{-\infty}^{\infty}\!\!\! {\rm d} v_2 \!\int_{-\infty}^{\infty} \!\!\!{\rm d} v_1 \, v_2 v_1 P(v_2,t_2 | v_1,t_1) W(v_1,t_1). \label{0}
\end{equation}
Here $P(v_2,t_2|v_1,t_1)$ is the conditional probability density, i.e. the probability to find a particle with velocity $v_2$ at time $t_2$, provided it had velocity $v_1$ at $t_1$, and $W(v_1,t_1)$ is  the probability density to find a particle with velocity $v_1$ at $t_1$. 
Equation \eqref{0} can be used to determine the mean-square displacement $\langle x^2(t)\rangle$, an  important transport characteristic. 
In the case  where  the process is stationary at long times, with stationary  density $W_{\text{s}}(v_1) = \lim_{t_1 \rightarrow \infty} W(v_1,t_1)$, and if $P(v_2,t_2 | v_1,t_1)= P(v_2,t_2-t_1 | v_1,0) $,  Eq.~\eqref{0} tends to the stationary correlation function $C_{v,{\rm s}}(\tau)=\langle v(\tau) v(0) \rangle_{\text{s}}$, which only depends on the  time lag $\tau  = t_2-t_1$,
\begin{equation}
C_{v,{\rm s}}(\tau) = \int_{-\infty}^{\infty} {\rm d} v_2 \int_{-\infty}^{\infty} {\rm d} v_1 \, v_2 v_1 P(v_2,\tau | v_1,0) W_{\text{s}}(v_1). \label{0a}
\end{equation}
Since   $x(t) = \int_{0}^{t} {\rm d}t_1 v(t_1)$, we can express the second moment of the position as $\langle x^2(t) \rangle = \int_{0}^{t} {\rm d}t_2 \int_{0}^{t} {\rm d}t_1 C_{v}(t_2,t_1)$ for a particle starting at the origin, $x(t)=0$ for $t=0$. 
The (one-dimensional) diffusion coefficient $D_1$ then follows as,
\begin{align}
D_1 = \frac{1}{2} \lim_{t \rightarrow \infty} \frac{{\rm d}}{{\rm d} t} \langle x^2(t) \rangle = \lim_{t \rightarrow \infty}  \int_{0}^{t} {\rm d} \tau \ C_{v,{\rm s}}(\tau), \label{4}
\end{align}
where the subscript $1$ indicates that diffusion is normal, $\langle x^2(t) \rangle \simeq 2 D_1 t^1$.
The Green-Kubo formula \eqref{4} only holds if the velocity correlation function $C_v(t_2,t_1)$ has a stationary limit and if it decays fast enough (e.g.~exponentially) so that  the time  integral  is finite.
This condition defines a certain class of correlation functions.
For this kind of system, even if the initial state is not the stationary one, it is justified to assume the validity of Eq.~\eqref{4}, as long as one is interested in the dynamics on time scales much longer than the characteristic correlation time.
In the following, we will introduce a different class of correlation functions, for which the Green-Kubo formula \eqref{4} does not hold.
Note that generally, there is also the possibility that the time integral in Eq.~\eqref{4} vanishes. This leads to subdiffusion, which we will not consider here.

We consider systems whose velocity correlation function has the following asymptotic scaling form when both $t$, called the age of the system, and $\tau$, the time lag, are large compared to the system's microscopic time scales,
\begin{align}
C_v(t+\tau,t) \simeq \mathcal{C} \ t^{\nu-2} \phi\left(\frac{\tau}{t}\right).  \label{6}
\end{align}
This expression is valid for $\tau > 0$; for $\tau < 0$, one has to replace $\tau$ by $-\tau$ and $t$ by $t+\tau$ in order to respect the symmetry of the correlation function with respect to interchanging its arguments.
Here $\mathcal{C}$ is a constant and $\nu > 1$.
The asymptotic behavior of the positive valued scaling function $\phi(s)$ is limited by power laws,
\begin{align}
\phi(s) < c_{\rm l} s^{-\delta_{\rm l}} \quad &\text{with} \quad 2-\nu \leq \delta_{\rm l} < 1 \quad \text{for} \quad s \rightarrow 0 \nonumber \\
\phi(s) < c_{\rm u} s^{-\delta_{\rm u}} \quad &\text{with} \quad \delta_{\rm u} > 1-\nu \quad \text{for} \quad s \rightarrow \infty,
\end{align}
where $c_{\rm l}$ and $c_{\rm u}$ are positive constants.
These conditions do not require $\phi(s)$ to behave as a power law at small and/or large arguments as long as the asymptotic behavior is within the above limits, so $\phi(s)$ might for example decay exponentially for large $s$.
Since the correlation function function Eq.~\eqref{6} explicitly depends on the age $t$ of the system, it is an aging correlation function, in contrast to a stationary correlation function which only depends on the time lag $\tau$.
The asymptotic mean square displacement $\langle x^2(t) \rangle$ for this type of correlation function is then given by,
\begin{align}
\langle x^2(t) \rangle \simeq 2 \mathcal{C} \int_{0}^{t} {\rm d} t_2 \int_{0}^{t_2} {\rm d} t_1 \ t_1^{\nu-2} \phi\left(\frac{t_2-t_1}{t_1}\right), \label{7}
\end{align}
where we have used the symmetry of the correlation function \eqref{6}.
Introducing the variable $s = (t_2-t_1)/t_1$, we obtain,
\begin{align}
\langle x^2(t) \rangle &\simeq 2 \mathcal{C} \int_{0}^{t} {\rm d} t_2 \, t_2^{\nu - 1} \int_{0}^{\infty} {\rm d}s \ (s+1)^{-\nu} \phi(s).
\end{align}
We thus arrive at our first main result,
\begin{align}
\langle x^2(t) \rangle &\simeq 2 D_{\nu} t^{\nu} \label{8} \\
\text{with} \quad D_{\nu} &= \frac{\mathcal{C}}{\nu} \int_{0}^{\infty} {\rm d} s \ (s+1)^{-\nu} \phi(s). \nonumber
\end{align}
While  the usual Green-Kubo relation \eqref{4} holds for normal diffusion, this scaling Green-Kubo relation \eqref{8} is applicable to $\nu > 1$, which corresponds to superdiffusion. 
Since then $\langle x^2(t) \rangle$ grows faster than linear in time, the usual diffusion coefficient in Eq.~\eqref{4} is not defined. 
Instead, we have the anomalous diffusion coefficient $D_{\nu}$, which, similarly to the original Green-Kubo formula \eqref{4}, is given in terms of an integral over a function of a single variable.
Determining the diffusive behavior of a system from its correlation function thus amounts to determining the exponent $\nu$ and the scaling function $\phi(s)$.

While we derived Eq.~\eqref{8} in terms of position and velocity, it holds for any two quantities where one is the time integral of the other. An example for this is given in Section \ref{SEC5}.

\subsection{Classification\label{SEC2A}}

The velocity correlation function Eq.~\eqref{6} includes two important special cases. The first are long range stationary correlation functions which exhibit a power law decay for large $\tau$,
\begin{align}
C_{v,{\rm s}}(\tau) \simeq \mathcal{C}_{\rm s} \ \tau^{\nu-2} \label{5a}
\end{align}
with $1 < \nu < 2$ and $\mathcal{C} > 0$ a constant. 
This is obtained from Eq.~\eqref{6} if the scaling function is given by $\phi(s) = s^{\nu-2}$.
While this correlation function is stationary, its decay is such that the time integral in Eq.~\eqref{4} diverges in the infinite time limit, so that the usual Green-Kubo formula is not applicable.
From the scaling Green-Kubo relation Eq.~\eqref{8}, we obtain,
\begin{align}
\langle x^2(t) \rangle &\simeq 2 D_{\nu,{\rm s}} t^{\nu} \quad \text{with} \quad D_{\nu,{\rm s}} = \frac{\mathcal{C}_{\rm s}}{\nu (\nu-1)} \label{GKstat} ,
\end{align}
where the subscript s stands for stationary.
In this case, the anomalous diffusion coefficient can also be expressed in terms of a fractional derivative of the stationary velocity autocorrelation function as was shown by Kneller \cite{kne11}. However, as we will see below, the stationary correlation function may not be sufficient to describe the long time diffusivity of the system.
The second special case is $\nu = 2$ where we have the usual type of aging correlation function, which is of the form \cite{bou92,cug94,bou95,bur10},
\begin{align}
C_v(t + \tau,t) \simeq \mathcal{C} \ \phi\left(\frac{\tau}{t}\right) . \label{5}
\end{align}
Since in this case, there exists no stationary correlation function, Eq.~\eqref{4} is not applicable.
For the aging type of correlation function, Eq.~\eqref{5}, we find ballistic behavior,
\begin{align}
\langle x^2(t) \rangle &\simeq 2 D_{2} t^2 \label{GKball} \\
\text{with} \quad D_{2} &= \frac{\mathcal{C}}{2} \int_{0}^{\infty} {\rm d} s \ (s+1)^{-2} \phi(s). \nonumber
\end{align}
For $\nu > 2$, we refer to the correlation function Eq.~\eqref{6} as superaging, since then, in addition to the dependence on the age of the system via the scaling function like in the usual aging case Eq.~\eqref{5}, the overall value of the correlation function also increases with the age of the system.
A similar type of correlation function, although with a logarithmic time dependence of the prefactor has been found for a random walker in a random environment \cite{dou99}.
Summarizing, we thus have four different classes of correlation functions: Stationary, integrable correlation functions are covered by the usual Green-Kubo formula Eq.~\eqref{4}. 
By contrast, stationary, nonintegrable correlation functions and aging correlation functions, which we can further differentiate into aging and superaging ones, are all encompassed by the scaling Green-Kubo relation Eq.~\eqref{8}.
Three important examples for physical models whose correlation functions belong to one of the latter three classes and thus follow the asymptotic behavior Eq.~\eqref{6} are discussed in section \ref{SECIII}.

The occurrence of the scaling function $\phi(s)$ in Eq.~\eqref{6} captures the fact that the age $t$ of the system also sets the time scale on which the correlations decay, i.e. the correlation time increases linearly with the age of the system. For the usual exponential kind of correlation functions, the correlation time is constant and provides a typical time scale. In aging systems, no constant correlation time exists or it is much larger than the time scales over which the system is observed. Then the only relevant time scale is the age of the system. If the correlation time increases slower or faster than linear with the age of the system, the diffusion may be retarded (and even subdiffusive) or accelerated compared to Eq.~\eqref{8}, depending on the asymptotic behavior of the scaling function $\phi(s)$. The case where the correlation time depends sublinearly on the age of the system, is also referred to as subaging \cite{rin01,mon03}. This retarded and accelerated aging is discussed in the appendix.

\begin{figure*}
\begin{center}
\includegraphics[trim=10mm 5mm 82mm 15mm, clip, width=0.245\textwidth]{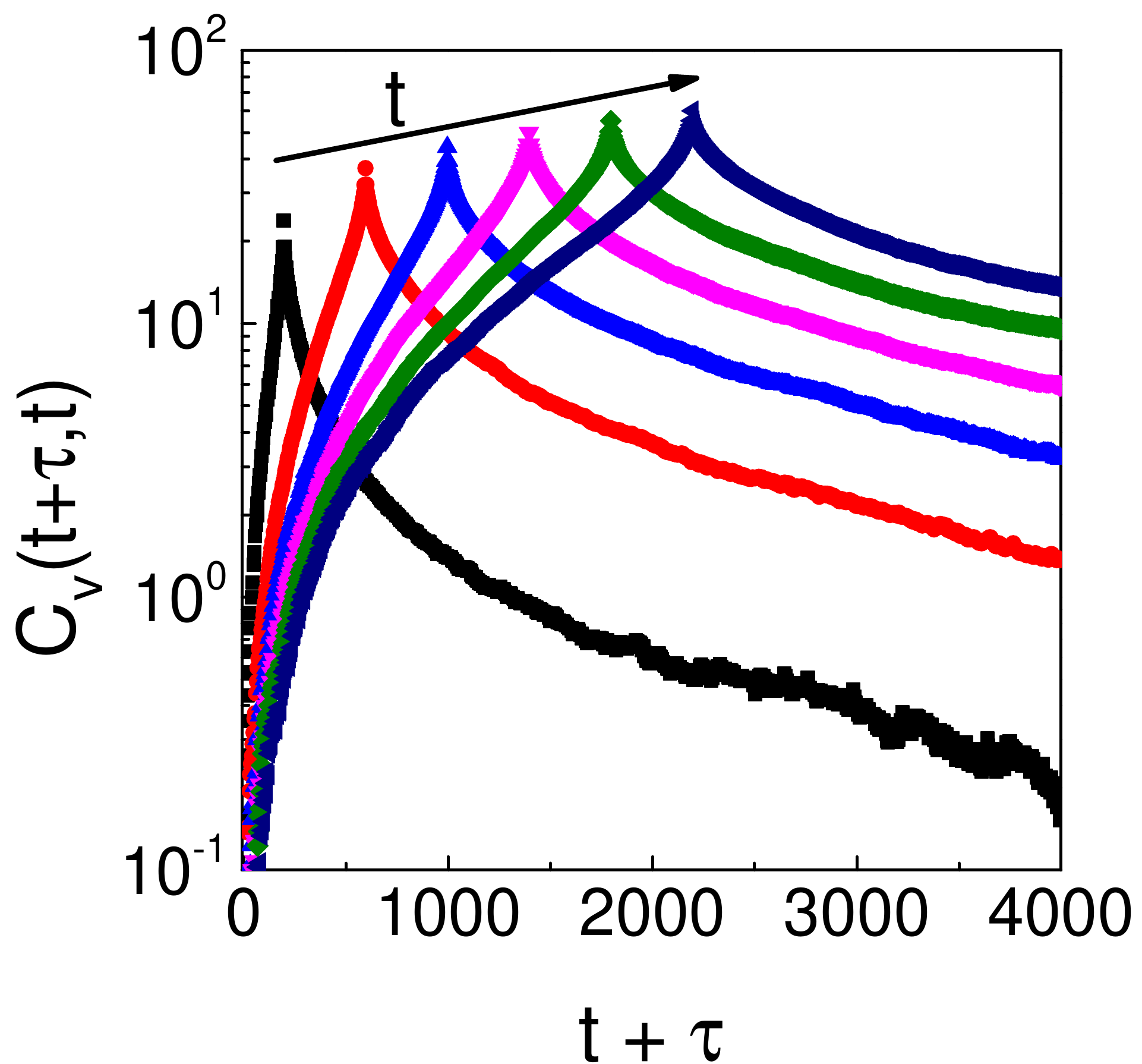}
\includegraphics[trim=10mm 5mm 82mm 15mm, clip, width=0.245\textwidth]{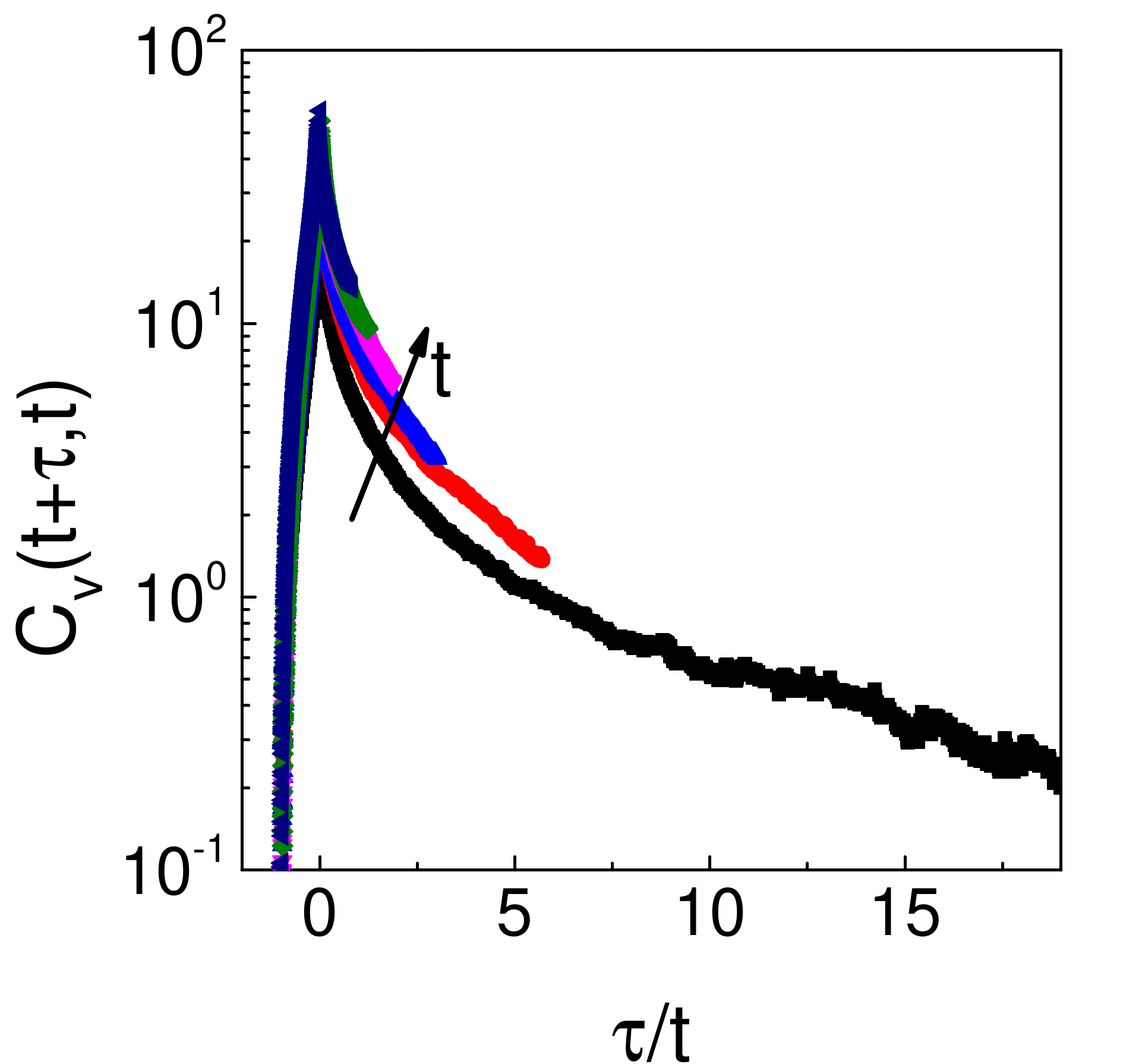}
\includegraphics[trim=10mm 5mm 82mm 15mm, clip, width=0.245\textwidth]{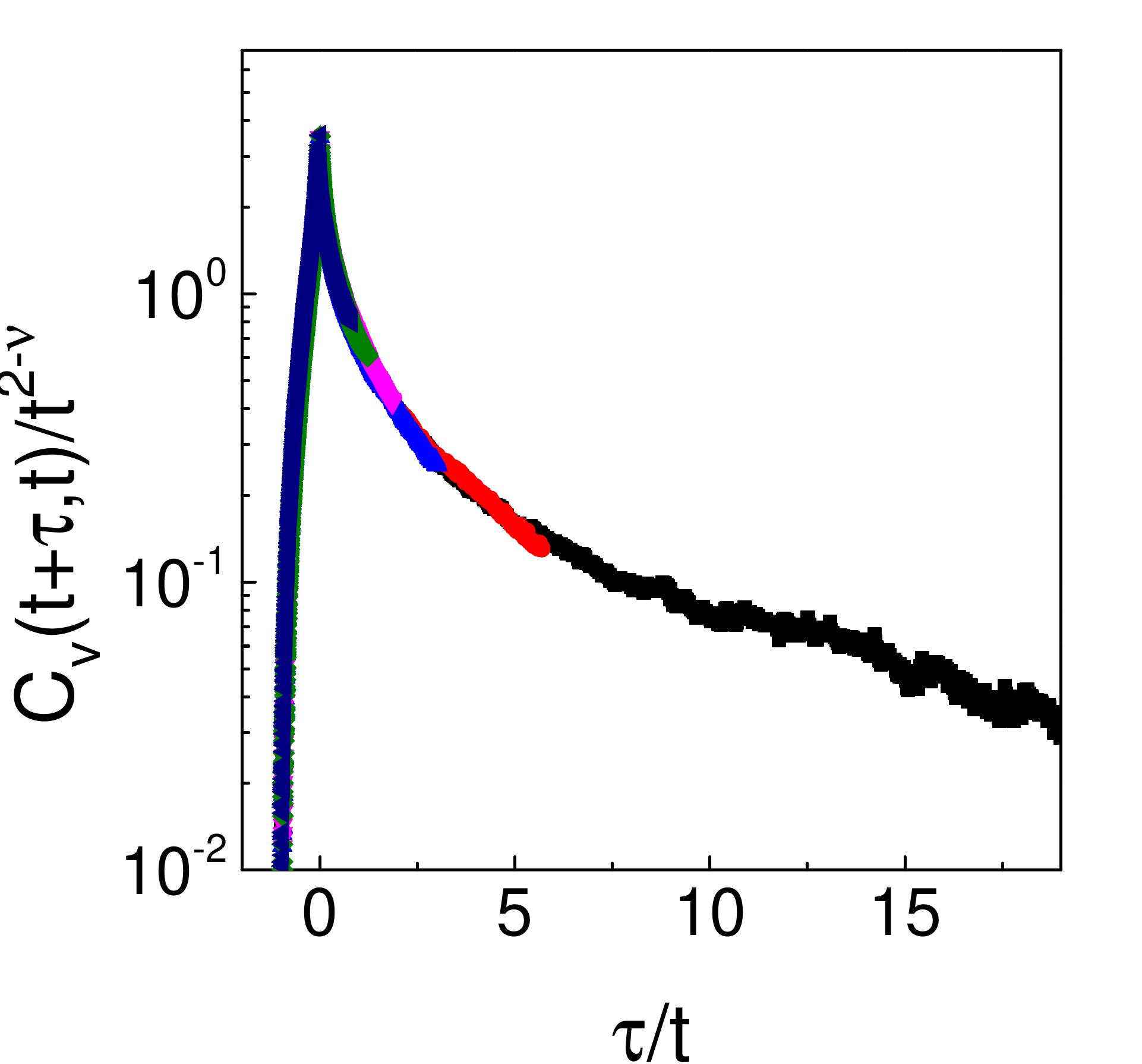}
\includegraphics[trim=10mm 5mm 82mm 15mm, clip, width=0.245\textwidth]{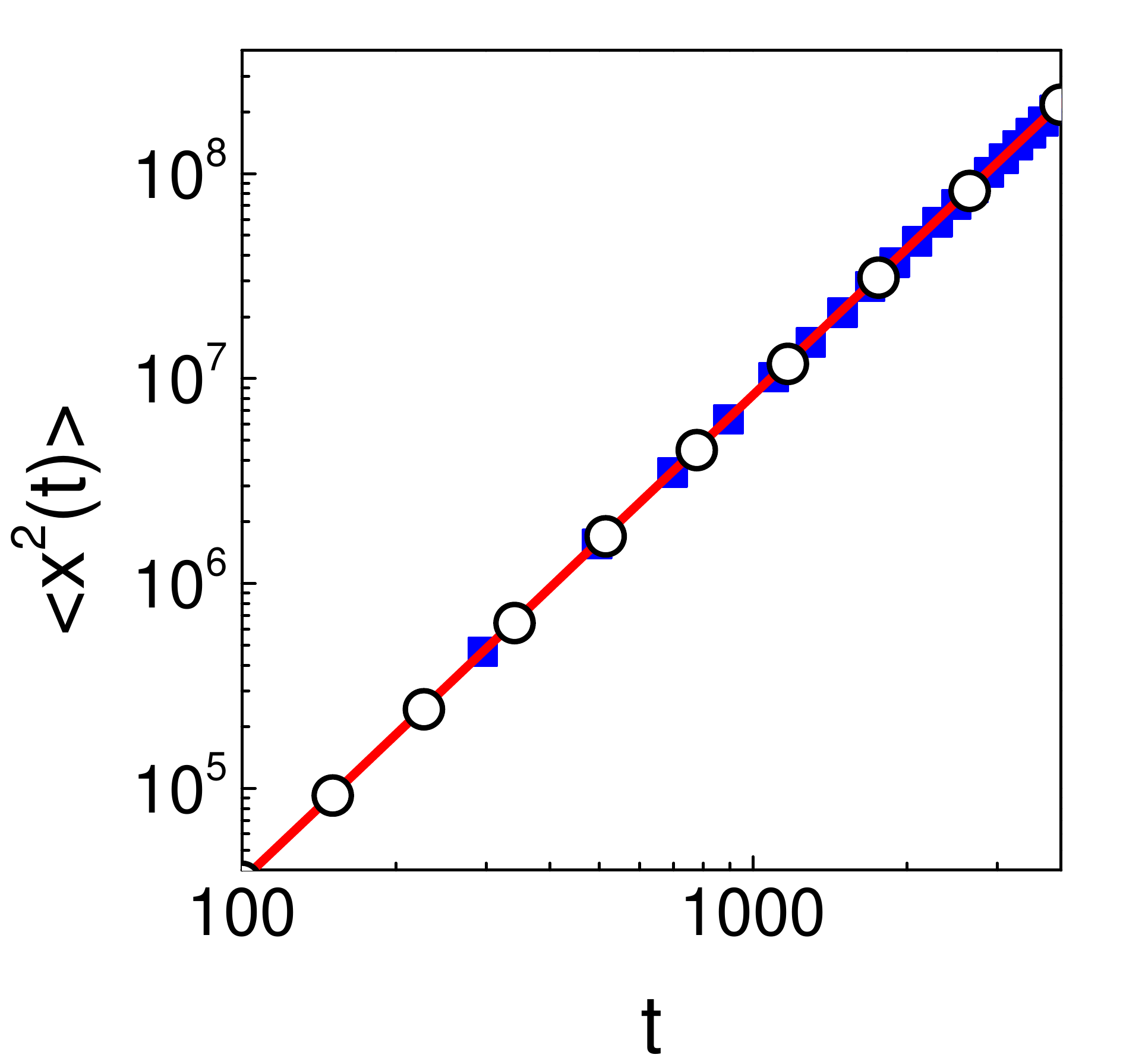}
\end{center}
\caption{Computation of the diffusion exponent and coefficient for cold atoms in optical lattices from the correlation function using the scaling Green-Kubo formula Eq.~\eqref{8}. a) Correlation function as a function of the total time $t+\tau$ for different values of $t$. b) Correlation function plotted as a function of $\tau/t$. c) After rescaling by $t^{\nu-2}$ with $\nu = 2.37$. d) Mean square displacement obtained by integrating over the rescaled correlation function $\phi(s)$ (empty circles) compared to the analytical result Eqs.~\eqref{lp_exp} and \eqref{lp_Dnu} (red line) and the mean square displacement obtained directly from the Langevin simulation (blue squares). From the rescaling and integration procedure, we obtain $\nu \simeq 2.37$ and $D_{\nu} \simeq 0.321$, which agrees well with the asymptotic analytical results $\nu = 2.37$ and $D_{\nu}=0.323$. The parameters for the simulation are $D = 3.25$, $\gamma = 5.10$, and $v_c = 1.20$ (see Eq.~\eqref{lp_FPE}) with $10^6$ trajectories and $2.5 \times 10^5$ integration steps per trajectory. \label{F4}} 
\end{figure*}

\subsection{Application to experimental data \label{SEC2C}}

Before we start discussing the application of the scaling Green-Kubo relation to specific physical systems, we outline its application as a data analysis method for a generic velocity correlation function that has been obtained experimentally or numerically.
To this end, we use the velocity correlation function generated by performing numerical Langevin simulations of free diffusion under the influence of a friction force that is inversely proportional to the velocity (see Section \ref{SEC3} for details).
For now, we do not discuss the specifics and properties of this system but instead just use it as a way to generate the data for the correlation function.
Applying Eq.~\eqref{8} to the measured correlation function requires identifying the scaling behavior of the latter. 
This is done in the following way, as illustrated in Fig.~\ref{F4}: First, we obtain the correlation function as a function of the time lag $\tau$ for different initial times $t$ and plot it as a function of $\tau/t$ (Fig.~\ref{F4}a) and b)). 
Then we rescale the resulting curves by $t^{\nu-2}$, the correct value of $\nu$ being obtained when the data collapses onto a single curve (Fig.~\ref{F4}c)). 
This determines the diffusion exponent $\nu$. The single rescaled curve is then the function $\phi(s)$ in Eq.~\eqref{8}. 
Multiplying $\phi(s)$ by $(1+s)^{-\nu}$ and numerically integrating the resulting curve, we obtain the generalized diffusion coefficient $D_{\nu}$. 
In Fig.~\ref{F4}d), we compare the results obtained in this way to the mean square displacement obtained directly from the numerical simulations by integrating the velocity process and also to the asymptotic analytical one (see Eqs.~\eqref{lp_exp} and \eqref{lp_Dnu}) and find good agreement within the accuracy of the numerical simulations.
Compared to determining the diffusion exponent and coefficient directly from the data for the mean square displacement, this rescaling method has the advantage that it does not require any fitting of the data.

\subsection{Persistence of initial conditions \label{SEC2B}}

Previously, we discussed the long time limit of the mean square displacement of a process starting at time $t=0$.
However, since the systems we are interested in exhibit long time correlations, the previous history of the system, i.e. the initial condition, may have a significant impact on the dynamics.
In order to discuss the effect of the initial condition, let us consider a slightly different situation. 
We still assume that the long time asymptotic form of the correlation function is given by Eq.~\eqref{6}. 
However, now consider the relative mean square displacement $\langle \Delta x^2(t) \rangle_{t_0} = \langle (x(t+t_0)-x(t_0))^2 \rangle$ with respect to some long aging time $t_0$.
We can think of $t_0$ as an "equilibration time": The process starts with a sharp initial condition (e.g. $v(0) = 0$, $x(0) = 0$) and we let it evolve for a time $t_0$ before starting to measure the mean square displacement. 
Note that generally we consider nonstationary systems, which possess no actual equilibrium state, hence the quotation marks on "equilibration time".
By definition, the relative mean square displacement is given by,
\begin{align}
\langle \Delta x^2(t) \rangle_{t_0} &= \langle (x(t_0+t) - x(t_0))^2 \rangle \nonumber \\
& = \int_{t_0}^{t_0+t} {\rm d}t_2 \int_{t_0}^{t_0+t} {\rm d}t_1 C_v(t_2,t_1) \nonumber \\
& \simeq 2 \mathcal{C} \int_{0}^{t} {\rm d} t_2 \int_{0}^{t_2} {\rm d} t_1 (t_1+t_0)^{\nu-2} \phi\left(\frac{t_2-t_1}{t_1+t_0}\right).
\end{align}
Introducing the variables $s = (t_2-t_1)/(t_1+t_0)$ and $z = t_2/t$, we can write this as,
\begin{align}
\langle \Delta x^2(t) \rangle_{t_0} &\simeq 2 D^{t/t_0}_{\nu} t^{\nu}  \nonumber \\
\text{with} \quad D_{\nu}^{t/t_0} &= \mathcal{C} \int_{0}^{1} {\rm d} z \ z^{\nu-1} \left(1+\frac{1}{z \frac{t}{t_0}} \right)^{\nu-1} \nonumber \\
& \qquad \times \int_{0}^{z \frac{t}{t_0}} {\rm d} s \ (s+1)^{-\nu} \phi(s). \label{GK_t0}
\end{align}
This resembles the scaling Green-Kubo relation \eqref{8}, however with an anomalous diffusion coefficient $D_{\nu}^{t/t_0}$ that formally depends on the ratio $t/t_0$. 
In the limit $t \gg t_0$, we can neglect the second term in parentheses and take the upper bound of the second integral in Eq.~\eqref{GK_t0} to infinity to obtain the expression given by the scaling Green-Kubo relation \eqref{8} independent of $t_0$,
\begin{align}
D_{\nu}^{\infty} = D_{\nu} = \frac{\mathcal{C}}{\nu} \int_{0}^{\infty} {\rm d} s \ (s+1)^{-\nu} \phi(s).
\end{align}
For finite equilibration times, the scaling Green-Kubo relation together with the aging correlation function Eq.~\eqref{6} thus gives the "true" long time limit for the mean square displacement.

In the opposite limit $t \ll t_0$, on the other hand, the behavior is governed by the small-$s$ expansion of the scaling function $\phi(s)$,
\begin{align}
D_{\nu}^{t/t_0} \simeq \mathcal{C} \left(\frac{t_0}{t}\right)^{\nu-1} \int_{0}^{1} {\rm d} z \int_{0}^{z \frac{t}{t_0}} {\rm d} s \ \phi(s).
\end{align}
Generally, if the second moment of the velocity is asymptotically either constant or increases with time, $\langle v^2(t) \rangle = C(t,t) \simeq a \mathcal{C} t^{\beta}$ with some positive constant $a$ and $0 \leq \beta < \nu - 1$, continuity demands $\phi(s) \simeq a s^{-\delta_{\rm l}}$ with $\delta_{\rm l} = 2-\nu+\beta$ for small $s$ and thus,
\begin{align}
D_{\nu}^{t/t_0} \simeq \frac{a \mathcal{C}}{(\nu-\beta+2)(\nu-\beta+1)} \left(\frac{t_0}{t}\right)^{\beta}.
\end{align}
Consequently, we find for the relative mean square displacement,
\begin{align}
\langle \Delta x^2(t) \rangle_{t_0} \simeq 2 \frac{a \mathcal{C}}{(\nu-\beta+2)(\nu-\beta+1)} t_0^{\beta} t^{\nu-\beta} \label{MSD_t0}.
\end{align}
This is obviously different from Eq.~\eqref{8}, in particular the diffusion exponent is different and the anomalous diffusion coefficient grows with the aging time $t_0$. 
This shows that the history of the process, and, by extension, the initial state of the systems at time $t_0$ influence the diffusive behavior. 
For $\beta > 0$, where the second moment of the velocity increases with time and thus no stationary correlation function exists, this is not surprising. If we let the process evolve for a longer time, the overall velocity will increase and thus diffusion becomes faster. 
Of particular interest is the case $\beta = \nu - 2 > 0$, that occurs for the superaging regime of the examples discussed in Sections \ref{SEC3} and \ref{SEC4}. 
Here we find from Eq.~\eqref{MSD_t0} quasi-ballistic diffusion with an anomalous diffusion coefficient that increases with $t_0$.

If the stationary correlation function Eq.~\eqref{5a} exists, the second moment of the velocity is asymptotically constant and thus $\beta = 0$. We then have with $\mathcal{C}_{\rm s} = a \mathcal{C}$,
\begin{align}
D_{\nu}^{0} = \frac{\mathcal{C}_{\rm s}}{\nu (\nu-1)} = D_{\nu,{\rm s}}.
\end{align}
This is precisely the stationary result Eq.~\eqref{GKstat}. In this case, we thus have two values for the anomalous diffusion coefficient,
\begin{align}
\langle \Delta x^2(t) \rangle_{t_0} \simeq \left\lbrace \begin{array}{ll}
2 D_{\nu} t^{\nu} &\text{for} \; t \gg t_0 \\[2 ex]
2 D_{\nu,{\rm s}} t^{\nu} &\text{for} \; t \ll t_0, 
\end{array} \right. \label{Dstat}
\end{align}
both of which are independent of $t_0$.
A particular case of Eq.~\eqref{Dstat} is $\nu = 2$, $\beta = 0$, when the aging correlation function Eq.~\eqref{5} tends to a constant which takes the place of the stationary correlation function for $t \gg \tau$.
An example of this is the aging correlation function for the L{\'e}vy walk, see Section \ref{SEC5}.
Which of the two values $D_{\nu}$ and $D_{\nu,{\rm s}}$ in Eq.~\eqref{Dstat} correctly describes the long time diffusive behavior depends on the initial preparation of the system: 
If the system is not initially in the stationary state (i.e. $t_0$ is finite), then we need to use the aging correlation function Eq.~\eqref{6} and obtain $D_{\nu}$ for the diffusivity. 
On the other hand, if the system is initially prepared in the stationary state (i.e. infinite $t_0$), then it is correctly described by the stationary correlation function Eq.~\eqref{5a} and we obtain $D_{\nu,{\rm s}}$ for the diffusivity. 
The initial condition of the system thus persists even in the long time limit. 
This is in contrast to systems where the correlations decay exponentially and the long time diffusivity is uniquely determined by the Green-Kubo formula \eqref{4} independent of the initial condition.
Examples for this persistence of the initial condition are discussed in Section \ref{SECIII}.

\section{Application to three scale invariant systems \label{SECIII}}

\subsection{Diffusion with a $1/v$ friction force \label{SEC3}}

Our first example for a scale invariant system to which our scaling Green-Kubo formula is applicable is diffusion with a friction force that is inversely proportional to the velocity. The velocity dynamics are governed by the Fokker-Planck equation,
\begin{align}
\partial_t W(v,t) &= \partial_v [-F(v) W(v,t)) + D \partial_v W(v,t)], \label{lp_FPE}
\end{align}
for the velocity probability density $W(v,t)$, $D$ being the velocity diffusion coefficient. We take the the friction force to be asymptotically inversely proportional to the velocity,
\begin{align}
F(v) \simeq - \frac{\gamma v_c^2 }{v} \quad \text{for} \quad v \gg v_c, \label{lp_friction_as}
\end{align}
with friction coefficient $\gamma$ and characteristic velocity $v_c$. A concrete example would be,
\begin{align}
F(v) = -\frac{\gamma v}{1+v^2/v_c^2}, \label{lp_friction}
\end{align}
which is linear for small velocities and has the desired behavior \eqref{lp_friction_as} for large velocities. This particular form of the friction force occurs in the semiclassical description of cold atoms in dissipative optical lattices \cite{cas90,mar96}. It turns out that the long time dynamics depend crucially on the ratio of the prefactor $\gamma v_c^2$ of the friction force and the diffusion coefficient $D$. Specifically we introduce the parameter,
\begin{align}
\alpha = \frac{\gamma v_c^2}{2 D} + \frac{1}{2}, \label{fp_alpha}
\end{align} 
in terms of which the stationary solution of Eq.~\eqref{lp_FPE} for $\alpha > 1$ is given by \cite{lut03},
\begin{align}
W_s(v) = \frac{1}{Z} \left( 1+  \frac{v^2}{v_c^2} \right)^{\frac{1}{2}-\alpha} \;  \text{with} \; 
Z = \frac{\sqrt{\pi}\Gamma(\alpha-1) v_c}{\Gamma\left(\alpha-1/2\right)} , \label{lp_stat}
\end{align}
where $\Gamma(a)$ denotes the Gamma function.
This stationary velocity distribution has power law tails, meaning that, depending on the exponent, different moments of the stationary distribution will be infinite. In particular, the second moment diverges when $\alpha < 2$, which implies infinite kinetic energy $E_k = m \langle v^2 \rangle /2$. For cold atoms in optical lattices, both the power law probability density and the divergence of the kinetic energy have been observed in experiments \cite{dou06,kat97}. Since an infinite kinetic energy is nonphysical, it is clear that a time dependent solution to the Fokker-Planck equation \eqref{lp_FPE} is required to capture the dynamics. Asymptotically for large velocities and long times, this time dependent solution is given by the infinite covariant density derived in \cite{kes10},
\begin{align}
W_\text{ICD}(v,t) \simeq \left\lbrace
\begin{array}{ll}
\frac{1}{Z \Gamma(\alpha)} \left(\frac{|v|}{v_c}\right)^{1-2\alpha} \Gamma \left( \alpha,\frac{v^2}{4 D t} \right) \\[1 ex]
\qquad \qquad \qquad \text{for} \; \alpha > 1 \\[2ex]
\frac{1}{\Gamma(1-\alpha)} (4 D t)^{\alpha - 1} |v|^{1-2\alpha} e^{-\frac{v^2}{4 D t}} \\[1 ex]
\qquad \qquad \qquad \text{for} \; \alpha < 1,
\end{array} \right. \label{lp_ICD} 
\end{align}
with the upper incomplete Gamma function $\Gamma(a,x)$.
In conjunction with the stationary solution, the infinite covariant density determines the asymptotic behavior of all moments, in particular,
\begin{align}
\langle v^2(t) \rangle \simeq \left\lbrace
\begin{array}{ll}
\frac{v_c^{2}}{2 (\alpha-2) }  \; &\text{for} \, \alpha > 2 \\[2ex]
\frac{v_c^{2\alpha-1}}{ Z \Gamma(\alpha) (2-\alpha)} (4 D t)^{2 -\alpha} \, &\text{for} \; 1 < \alpha < 2 \\[2ex] 
(1-\alpha) 4 D t \, &\text{for} \; \alpha < 1.
\end{array} \right. \label{lp_p2}
\end{align}
For $\alpha < 2$, where the second moment of the stationary distribution is infinite, the kinetic energy thus increases with time.
The corresponding velocity correlation function was derived in \cite{dec11,dec12a} and has the asymptotic behavior,
\begin{align}
C_v(t+&\tau,t) \simeq \nonumber \\
& \left\lbrace 
\begin{array}{ll}
v_c^{2 \alpha - 1}\frac{\sqrt{\pi}}{Z \Gamma(\alpha) \Gamma(\alpha+1)} \, (4 D t)^{2-\alpha} f_{\alpha } \left( \frac{\tau}{t} \right) \\ 
\qquad \text{for} \; \alpha > 1 \\[2ex]
\frac{\pi}{\Gamma(1 - \alpha) \Gamma(\alpha+1)} \, 4 D t \, g_{\alpha} \left( \frac{\tau}{t} \right) \\
\qquad \text{for} \; \alpha < 1,
\end{array} \right.  \label{lp_corr}
\end{align}
with the functions $f_{\alpha}(s)$ and $g_{\alpha}(s)$ defined by,
\begin{align}
f_{\alpha}(s) &= s^{2-\alpha} \int_{0}^{\infty} {\rm d} y \ y^{2} e^{-y^2} \ {}_{1}\text{F}_{1} \left( \frac{3}{2}; \alpha + 1; y^2 \right) \nonumber \\
& \qquad \qquad \qquad \qquad \times \Gamma \left( \alpha, y^2 s \right) , \nonumber \\
g_{\alpha}(s) &=  s \int_{0}^{\infty} {\rm d} y \ y^{2} e^{-y^2} \ {}_{1}\text{F}_{1} \left( \frac{3}{2}; \alpha + 1; y^2 \right) e^{-y^2 s},\label{lp_fg}
\end{align}
where ${}_{1}\text{F}_{1}(a;b;x)$ is a hypergeometric function.

For $\alpha < 3$, the correlation function Eq.~\eqref{lp_corr} is precisely of the type \eqref{6}, and we can apply our scaling Green-Kubo relation \eqref{8} and immediately obtain for the diffusion exponent,
\begin{align}
\nu = \left\lbrace \begin{array}{ll}
4-\alpha &\text{for} \quad 1 < \alpha < 3\\
3 &\text{for} \quad \alpha < 1. \label{lp_exp}
\end{array} \right.
\end{align}
Similarly, the anomalous diffusion coefficient is given by,
\begin{align}
\frac{D_{\nu} \theta^{\nu}}{\xi^{2} } & = \left\lbrace \begin{array}{ll}
\frac{4^{2-\alpha} \Gamma\left(\alpha-\frac{1}{2}\right)}{\Gamma(\alpha+1) \Gamma(\alpha) \Gamma(\alpha - 1)(4 - \alpha)} \int_{0}^{\infty} {\rm d} s (s+1)^{\alpha-4} f_{\alpha}(s) \\[1 ex]
\qquad \qquad \qquad  \qquad \qquad \qquad \text{for} \; 1 < \alpha < 3 \\[2 ex]
\frac{4 \sqrt{\pi}}{3 \Gamma(\alpha) \Gamma(1-\alpha)} \int_{0}^{\infty} {\rm d} s (s+1)^{-3} g_{\alpha}(s) \\[1 ex]
\qquad \qquad \qquad \qquad \qquad \qquad \text{for} \; \alpha < 1,
\end{array} \right. \label{lp_Dnu}
\end{align}
with the time and length scales $\theta = v_c^2/D$ and $\xi = v_c^3/D$. 
For $\alpha > 3$, the usual Green-Kubo formula \eqref{4} holds.
We can thus identify three different diffusion phases \cite{bar12}: normal diffusion $\langle x^2(t)\rangle \sim t$ when $\alpha>3$, anomalous superdiffusion $\langle x^2(t)\rangle \sim t^{4-\alpha}$ when $1<\alpha<3$, and Richardson-like diffusion $\langle x^2(t)\rangle \sim t^3$ when $\alpha<1$, which is known from turbulence \cite{fal01} and has recently been observed as a transient behavior in ordinary Brownian motion \cite{dup13}. 

\begin{figure}
\begin{center}
\includegraphics[trim=20mm 8mm 30mm 10mm, clip, width=0.45\textwidth]{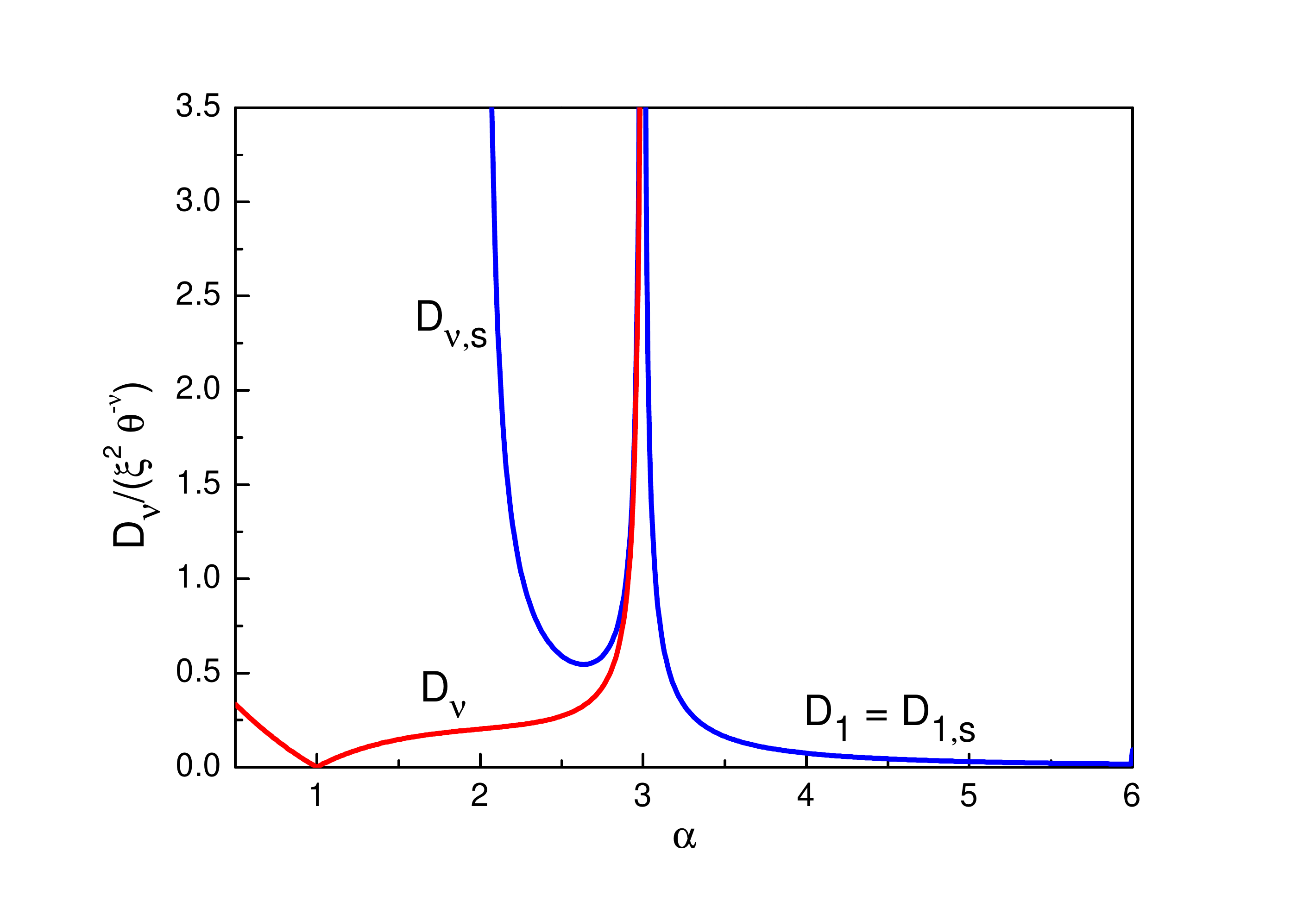}
\caption{Diffusion coefficient $D_{\nu}/(\xi^{2} \theta^{-\nu})$ \eqref{lp_Dnu} for cold atoms in optical lattices as a function of the parameter $\alpha$ (red line); the result for the normal diffusive case $\alpha > 3$ was taken from \cite{per00}. The result $D_{\nu, {\rm s}}$ from using the stationary correlation function Eq.~\eqref{lp_Dstat} is shown in blue. The coefficient $D_{\nu}$ diverges at the transition from normal to anomalous superdiffusion ($\alpha = 3$) and vanishes at the transition from superdiffusion to $t^3$-diffusion ($\alpha = 1$). While the values obtained from the aging and stationary correlation functions agree if the system is in the normal phase $\alpha > 3$, they differ significantly as the sytem becomes more superdiffusive, clearly signifying the dependence on the initial condition. As $\alpha$ approaches the transiton to aging behavior at $\alpha = 2$, the stationary correlation function ceases to provide a good description of the finite time system and the corresponding diffusion coefficient diverges. \label{F2}}
\end{center}
\end{figure}

As the age of the systems tends to infinity, $t \gg \tau$, the velocity correlation function Eq.~\eqref{lp_corr} attains a stationary limit for $\alpha > 2$ \cite{mar96},
\begin{align}
C_{v, {\rm s}}(\tau) \simeq v_c^{2 \alpha - 1} \frac{\pi \, \Gamma(\alpha - 2) }{4 Z \, \Gamma^2 \left( \alpha - \frac{1}{2} \right) } (4 D \tau)^{2-\alpha}. \label{lp_corrstat}
\end{align}
Applying the scaling Green-Kubo relation Eq.~\eqref{8} to the stationary correlation function, Eq.~\eqref{lp_corrstat}, we find for $2 < \alpha < 3$ by virtue of Eq.~\eqref{GKstat},
\begin{align}
\frac{D_{\nu,{\rm s}}}{\xi^{2} \theta^{-\nu}} = \frac{4^{1-\alpha} \sqrt{\pi}}{\Gamma\left(\alpha-\frac{1}{2}\right)(\alpha-2)(4-\alpha)(3-\alpha)}. \label{lp_Dstat}
\end{align}
Both the stationary and nonstationary expression the anomalous diffusion coefficient are plotted in Fig.~\ref{F2}.
Clearly, the two results differ substantially as the system approaches the aging phase $\alpha < 2$.
Both results describe the asymptotic long time behavior, but correspond to different initial conditions.
For the stationary expression $D_{\nu,{\rm s}}$, Eq.~\eqref{lp_Dstat}, the initial velocity probability distribution is the stationary power law distribution, Eq.~\eqref{lp_stat}.
By contrast, if the system starts out with a narrow (e.g. Gaussian) velocity initial condition, the asymptotic mean square displacement is described by $D_{\nu}$, Eq.~\eqref{lp_Dnu}.
In Refs.~\cite{hir11,hir12} it was observed that these two classes of initial conditions lead to a qualitatively different time evolution of the velocity probability density.
Our scaling Green-Kubo relation shows that for $2 < \alpha < 3$, the initial condition also persists in the diffusive behavior of the system in the form of different expressions for the anomalous diffusion coefficient.
For $\alpha < 2$, there is no stationary velocity correlation function and the system ages. 
Here, Eqs.~\eqref{lp_exp} and \eqref{lp_Dnu} describe the diffusive behavior on time scales that are much longer than the initial relaxation time of the system (see also Eq.~\eqref{GK_t0}).
For long relaxation times, the diffusion becomes quasi-ballistic, as was discussed in Ref.~\cite{dec12}.

The diffusion coefficient $D_{\nu}$ \eqref{lp_Dnu} can be seen (Fig.~\ref{F2}) to diverge at the transition from normal to superdiffusion ($\alpha = 3$), and in contrast, to vanish at the transition from superdiffusion to $t^3$-diffusion ($\alpha = 1$). 
This irregular behavior of the diffusion coefficient at the transitions justifies referring to the three different regimes as diffusion phases, as it indicates a change in the qualitative dynamics of the system.

\subsection{Fractional Langevin equation \label{SEC4}}

Next, we will consider the motion of a tracer particle in a crowded environment under the influence of external noise. The motion is governed by the fractional Langevin equation \cite{man68,lut01,pot03,goy07,bur08,bru09,tal10,liz10,sok12},
\begin{align}
\dot{v}(t) = -\int_{0}^{t} {\rm d} t' \gamma(t-t') v(t') + \frac{1}{m}\eta(t), \label{GLE}
\end{align}
where the memory kernel $\gamma(t)$ describes the retarding friction experienced by the tracer in a crowded environment \cite{wei13}, like the cytoplasm of biological cells. 
This friction kernel is given by,
\begin{align}
\gamma(t) = \frac{\gamma_{\rho}}{\Gamma(\rho)} t^{\rho-1} \label{friction},
\end{align}
with $0 < \rho < 1$ and a generalized friction coefficient $\gamma_{\rho}$. 
If the noise $\eta(t)$ is thermal equilibrium noise, it is related to the friction kernel via the fluctuation-dissipation theorem,
\begin{align}
\langle \eta(t_2) \eta(t_1) \rangle = m k_B T \gamma(t_2-t_1), \label{FDT}
\end{align}
which expresses the fact that both friction and internal noise are due to interaction with the thermal bath. In this case, Eqs.~\eqref{GLE} and \eqref{friction} describe the motion of the tracer in a crowded equilibrium environment, which is known to exhibit subdiffusion with exponent $\nu = 1-\rho$ \cite{lut01,pot03}.
If $\eta(t)$ is considered to be an external noise driving the system out of equilibrium, the fluctuation-dissipation theorem does not hold, and the system may become superdiffusive \cite{bru09,dec13}.
In living cells, this external noise is due to the action of molecular motors and drives active transport \cite{bru09}. It requires an external source of energy (e.g. ATP) to continuously keep the system out of equilibrium \cite{gal09}.

Like the equilibrium noise, we take the external noise to be a Gaussian, power law correlated process. However, we allow the external noise to be nonstationary, which, as shown in \cite{dec13}, is necessary to explain the superdiffusion exponent observed in several experiments \cite{tre08,gal09}. More precisely, we define the autocorrelation of the noise in terms of its Laplace transform $\tilde{\eta}(s) = \int_{0}^{\infty} {\rm d} t e^{-s t} \eta(t)$ \cite{dec13},
\begin{align}
\langle \tilde{\eta}(s_2) \tilde{\eta}(s_1) \rangle = \mathcal{F}_{\lambda} \frac{(s_2 s_1)^{-\frac{\lambda}{2}}}{s_2 + s_1} \label{noise},
\end{align}
with $\lambda > \rho$ and the constant $\mathcal{F}_{\lambda}$ describing the magnitude of the external noise.
The condition $\lambda > \rho$ ensures that the external noise correlations decay more slowly than those of the internal noise. 
In the time domain, the noise correlations read,
\begin{align}
\langle \eta(t+\tau) \eta(t) \rangle = (-1)^{-\frac{\lambda}{2}} \frac{\mathcal{F}_{\lambda}}{\Gamma^2\left(\frac{\lambda}{2}\right)} \tau^{\lambda-1} {\rm B} \left(- \frac{t}{\tau}, \frac{\lambda}{2}, \frac{\lambda}{2} \right),
\end{align}
where ${\rm B}(x,a,b)$ is the incomplete Beta function.
The noise process has a stationary autocorrelation for $\lambda < 1$ in the limit $t \rightarrow \infty$, specifically $\langle \eta(t+\tau) \eta(t) \rangle \propto \tau^{\lambda-1}$.
For $\lambda > 1$ on the other hand, the noise correlations are nonstationary, here the variance of the external noise increases with time as $\langle \eta^2(t) \rangle \propto t^{\lambda-1}$. 
\footnote{For $\lambda < 1$, the noise process corresponds to fractional Gaussian noise in the long time limit, whereas for $\lambda > 1$, it is identical to Riemann-Liouville fractional Brownian motion with Hurst exponent $H = (\lambda-1)/2$.}

\begin{figure}
\begin{center}
\includegraphics[trim=15mm 6mm 25mm 10mm, clip, width=0.45\textwidth]{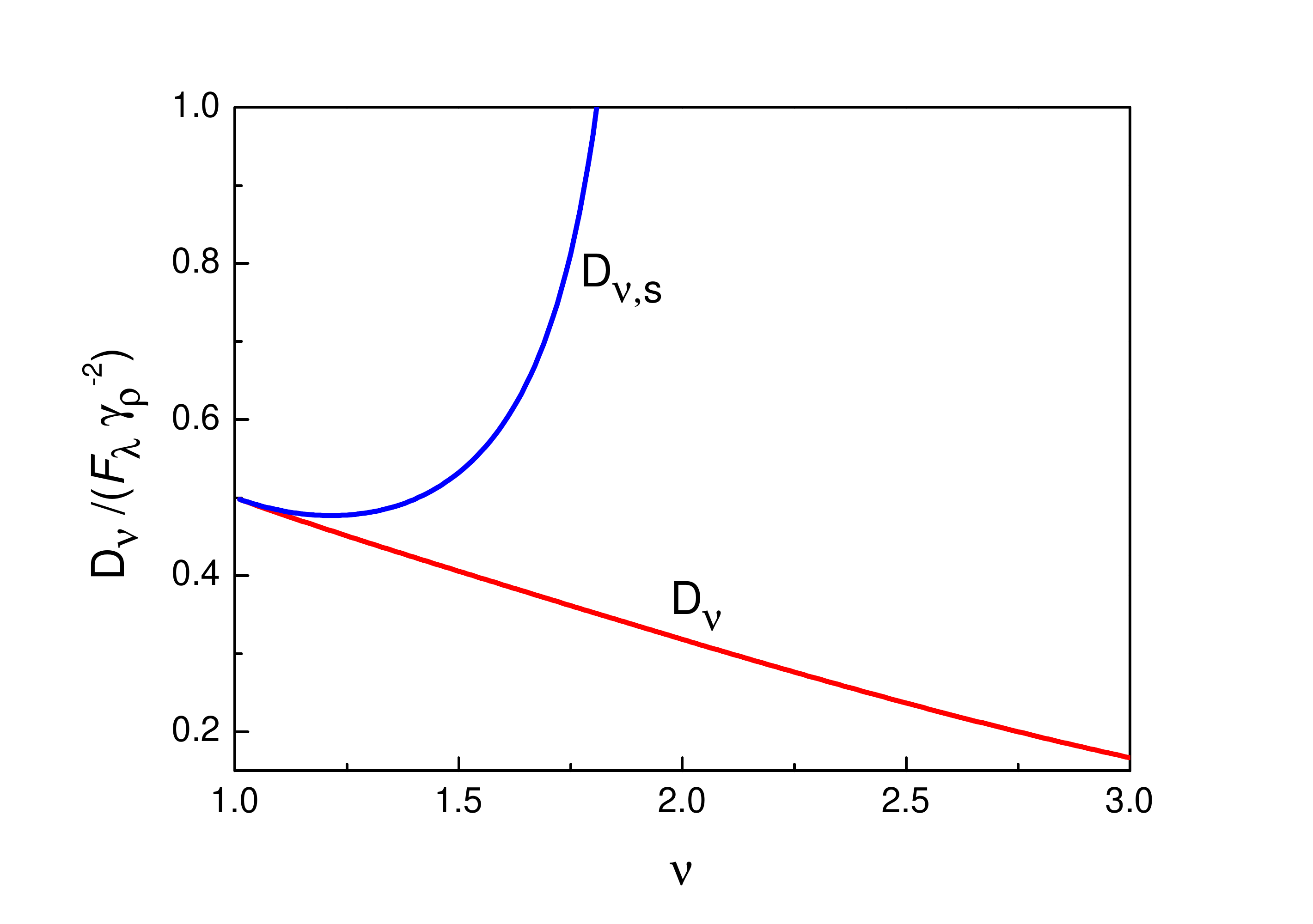}
\caption{The anomalous diffusion coefficient $D_{\nu}/(\mathcal{F}_{\lambda} \gamma_{\rho}^{-2})$ for the fractional Langevin equation as a function of the diffusion exponent $\nu$. The red line is the result using the aging correlation function Eq.~\eqref{GK_FBM} and the blue line the one from the stationary correlation function Eq.~\eqref{FLEstat}. Close to the threshold to normal diffusion $\nu = 1$, both values agree, however as the diffusion exponent increases the two results deviate from each other. At the transition to the nonstationary phase ($\nu = 2$), the stationary result diverges.\label{F8}}
\end{center}
\end{figure}

The velocity autocorrelation can be obtained by solving Eq.~\eqref{GLE} in Laplace space and then inverting the Laplace transform. 
The exact solution can be expressed in terms of the generalized Mittag-Leffler function $E_{a,b}(x)$ \cite{dec13},
\begin{align}
C_v(t+\tau,t) &= \frac{\mathcal{F}_{\lambda}}{m^2} \int_{0}^{t} {\rm d} t' G_{\rho,\lambda}( \tau + t') G_{\rho,\lambda}(t') \label{pcorrex} \\
\text{with} \quad G_{\rho,\lambda}(t) &= t^{\frac{\lambda}{2}} E_{\rho+1,\frac{\lambda}{2}+1}(-\gamma_{\rho} t^{\rho+1}) \nonumber .
\end{align}
The scaling properties of this expression become apparent when considering the limit of long times $\tau, t \gg \gamma_{\rho}^{-1/(1+\rho)}$,
\begin{align}
C_v(t+&\tau,t) \simeq \nonumber \\
&\frac{\mathcal{F}_{\lambda}}{m^2 \gamma_{\rho}^2 \Gamma\left(\frac{\lambda}{2}-\rho\right)\Gamma\left(\frac{\lambda}{2}-\rho+1\right)} \bigg[ (t+\tau)^{\frac{\lambda}{2}-\rho-1} t^{\frac{\lambda}{2}-\rho} \nonumber \\
&\; + (-1)^{\rho-\frac{\lambda}{2}} \left(\frac{\lambda}{2}-\rho-1\right) \tau^{\lambda-2\rho-1} \nonumber \\
&\; \; \times {\rm B} \left(-\frac{t}{\tau},\frac{\lambda}{2}-\rho+1,\frac{\lambda}{2}-\rho-1\right) \bigg]. \label{pcorr}
\end{align}
Casting this in the form of Eq.~\eqref{6}, we have,
\begin{align}
C_v&(t+\tau,t) \simeq \mathcal{C} \ t^{\nu-2} \phi\left(\frac{\tau}{t}\right) \label{GK_FBM} \\
&\text{with} \quad \nu = \lambda-2\rho+1 , \quad \mathcal{C} = \frac{\mathcal{F}_{\lambda}}{m^2 \gamma_{\rho}^2 \Gamma\left(\frac{\nu-1}{2}\right)\Gamma\left(\frac{\nu+1}{2}\right)} \nonumber \\
&\text{and} \quad \phi(s) = (1+s)^{\frac{\nu-3}{2}} \nonumber \\
&\; + (-1)^{\frac{1-\nu}{2}} \frac{\nu-3}{2} s^{\nu-2} {\rm B}  \left(-\frac{1}{s},\frac{\nu+1}{2},\frac{\nu-3}{2}\right)  . \nonumber
\end{align}
We can immediately read off the diffusion exponent $\nu=\lambda-2\rho+1$ and find that the system is superdiffusive for $\lambda > 2 \rho$.
Using our scaling Green-Kubo relation \eqref{8}, we can then calculate the anomalous diffusion coefficient by integrating the scaling function $\phi(s)$,
\begin{align}
D_{\nu} = \frac{\mathcal{F}_{\lambda}}{m^2 \gamma_{\rho}^2} \frac{1}{2 \nu \Gamma^2\left(\frac{\nu+1}{2}\right)} . \label{FLE_Dnu}
\end{align}
In the regime $1 < \nu < 2$, that is, for subballistic superdiffusion, the velocity autocorrelation Eq.~\eqref{pcorr} attains a stationary limit,
\begin{align}
C_{v,{\rm s}}(\tau) \simeq \frac{\mathcal{F}_{\lambda}}{\pi m^2 \gamma_{\rho}^2} \Gamma(2-\nu) \sin\left(\pi \frac{\nu-1}{2} \right) \tau^{\nu-2}. \label{FLEstat}
\end{align}
Similarly to Section \ref{SEC3}, we can apply the scaling Green-Kubo relation Eq.~\eqref{8} to both Eqs.~\eqref{GK_FBM} and \eqref{FLEstat}. From the latter, we obtain,
\begin{align}
D_{\nu,{\rm s}} = \frac{\mathcal{F}_{\lambda}}{m^2 \gamma_{\rho}^2} \frac{\Gamma(2-\nu) \sin\left(\pi \frac{\nu-1}{2} \right)}{\pi \nu (\nu-1)} . \label{FLE_Dnu_stat}
\end{align}
The results for the anomalous diffusion coefficient are shown in Fig.~\ref{F8}.
Once again, we observe significant differences between using the stationary and aging correlation functions as the diffusion exponent increases.
In contrast to the system discussed in section \ref{SEC3}, the stationary state is not fully described by the corresponding velocity distribution. 
Since the fractional Langevin equation \eqref{GLE} is non-Markovian, we need to specify the entire history of the process to determine the correct stationary state. 
Even if the velocity is initially distributed according to the stationary distribution, which is Gaussian with width $\langle v^2 \rangle_{\rm s} = (\mathcal{F}_{\lambda}/m^2) \int_{0}^{\infty} {\rm d} t \ G^2_{\rho, \lambda}(t)$, we still obtain the nonstationary result Eq.~\eqref{FLE_Dnu} for the diffusion coefficient. 
The stationary result Eq.~\eqref{FLE_Dnu_stat} is only obtained if we explicitly let the process evolve for some long time $t_0$ before starting the measurement of the mean square displacement.

In the regime $\nu > 2$, the velocity autocorrelation Eq.~\eqref{pcorr} is superaging and the variance of the velocity increases with time. The results for the diffusion exponent $\nu = \lambda - 2 \rho + 1$ and the diffusion coefficient $D_{\nu}$, Eq.~\eqref{FLE_Dnu}, remain unchanged. However, similarly to the superaging case in Section \ref{SEC3}, we then find faster than ballistic diffusion in this regime.

\subsection{Blinking quantum dots and L{\'e}vy walk \label{SEC5}}

As our final example for the application of the scaling Green-Kubo relation, we study the time dependent intensity $I(t)$ of the light emitted by nanocrystals embedded in a disordered medium. 
This deviates from the discussion so far in that the observables of interest are no longer velocity and position. Instead, we identify the velocity with the intensity and the total number of emitted photons, which is proportional to the time integral over the intensity, as the position.
The intensity exhibits a phenomenon called blinking: The quantum dot will emit light during some periods and remain dark during others \cite{kun00,kun01,shi01}. 
The lengths of these on and off periods are found to be distributed according to a power law whose exponent is such that the average on and off time is infinite (i.e. of the order of the measurement time), leading to aging in the intensity autocorrelation function \cite{bro03,mar04}.
This model is the familiar L{\'e}vy walk \cite{shl87,kla90,zum93,kla96} which is a popular stochastic framework with many applications.
As mentioned in the introduction, Zumofen and Klafter \cite{zum93} showed that the diffusivity in this model is sensitive to the initial conditions.
Here we demonstrate the applicability of our scaling Green-Kubo relation, while previous works used renewal theory approaches \cite{god01,mar04}.

\begin{figure}
\begin{center}
\includegraphics[trim=10mm 6mm 25mm 10mm, clip, width=0.45\textwidth]{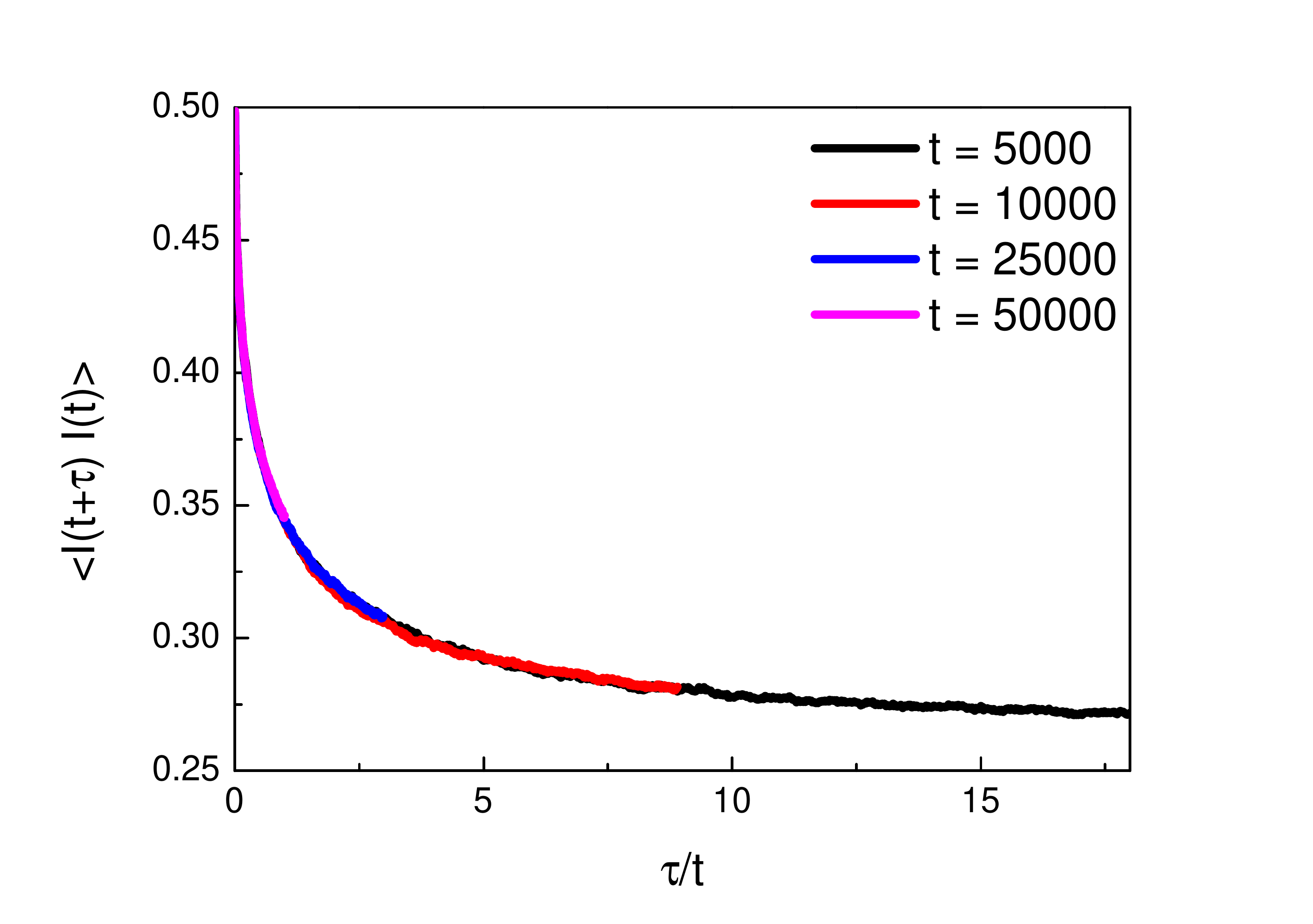}
\caption{Aging correlation function for the blinking quantum dots. The plots were obtained from numerical L{\'e}vy walk simulations with $t_{\rm c} = 1$ and $\mu = 0.6$. Since the correlation function is of the usual aging type ($\nu = 2$), no rescaling is necessary for the curves to collapse onto a single one. Integrating this curve, we obtain the coefficient $D_2 \simeq 0.043$ in good agreement with the result $D_2 = 0.05$ from Eq.~\eqref{MSD_BQD}. \label{F7}}
\end{center}
\end{figure}

We consider the special case where the on ($+$) and off ($-$) times of a blinking quantum dot follow the same power law distribution $\psi_{+}(t) = \psi_{-}(t) = \psi(t) \sim (t/t_{\rm c})^{-\mu-1}$ for $t \gg t_{\rm c}$ with $0 < \mu < 2$.
Thus the dot is in the on state with intensity $I=1$ for a waiting time drawn from $\psi(t)$, after which it switches to the off state $I=0$, again with a sojourn time drawn from $\psi(t)$ and so on.
The average on/off time $\bar{t}$ only exists for $\mu > 1$ and is infinite for $0<\mu<1$.
For this form of the distribution, the aging intensity autocorrelation for long times and $0 < \mu < 1$ has been explicitly obtained in Refs.~\cite{god01,mar04} by mapping the problem onto a L{\'e}vy walk,
\begin{align}
\langle &I(t+\tau) I(t) \rangle \nonumber \\ 
&\simeq \frac{1}{2} \left(1 - \frac{\sin(\pi \mu)}{2 \pi} {\rm B}\left(\frac{1}{1+\frac{t}{\tau}}, 1-\mu, \mu \right) \right) \label{icorr}.
\end{align}
Comparing this to the shape of the correlation function Eq.~\eqref{6}, we find $\nu = 2$, $\mathcal{C} = 1/2$ and,
\begin{align}
\phi(s) = 1 - \frac{\sin(\pi \mu)}{2 \pi} {\rm B}\left(\frac{s}{1+s}, 1-\mu, \mu \right).
\end{align}
Figure \ref{F7} shows the correlation function Eq.~\eqref{icorr} for $\mu = 0.6$ obtained from numerical simulations. 
Plotting the intensity autocorrelation as a function of $\tau/t$ results in a data collapse without any rescaling, indicating the pure aging behavior for $0 < \mu < 1$.
The total number of photons $n(t)$ emitted by the quantum dot during time $t$ is proportional to the time integral of the intensity, $n(t) = \mathcal{I}_0 \int_{0}^{t} {\rm d} t' I(t')$ with $\mathcal{I}_0$ the physical intensity of the dot in the on state.
Consequently, $\nu = 2$ in the scaling Green-Kubo relation Eq.~\eqref{8} immediately implies that the second moment of the photon number exhibits ballistic scaling (see Eq.~\eqref{GKball}) \cite{kla90}:
\begin{align}
\langle n^2(t) \rangle &\simeq \frac{\mathcal{I}_0^2}{2}\left(1-\frac{\mu}{2}\right) t^2 \label{MSD_BQD} .
\end{align}
Since the average intensity is asymptotically constant, $\langle I(t) \rangle \simeq 1/2$, we find for the variance of the photon number,
\begin{align}
\langle\delta n^2(t) \rangle &= \langle n^2(t) \rangle - \langle n(t) \rangle^2 \simeq 2 D_2 t^2 \nonumber \\
\text{with} \quad D_2 &= \frac{\mathcal{I}_0^2}{8} (1-\mu), \label{photvar_ball}
\end{align}
which also scales ballistically. This agrees with the result obtained in Ref.~\cite{jun01} using the probability distribution $P(n,t)$ of the photon number and with the experimental observation of  a linear increase of Mandel's $Q$-parameter $Q = \langle \delta n^2(t) \rangle/\langle n(t) \rangle$ \cite{mar06}.

\begin{figure}
\begin{center}
\includegraphics[trim=15mm 6mm 25mm 10mm, clip, width=0.45\textwidth]{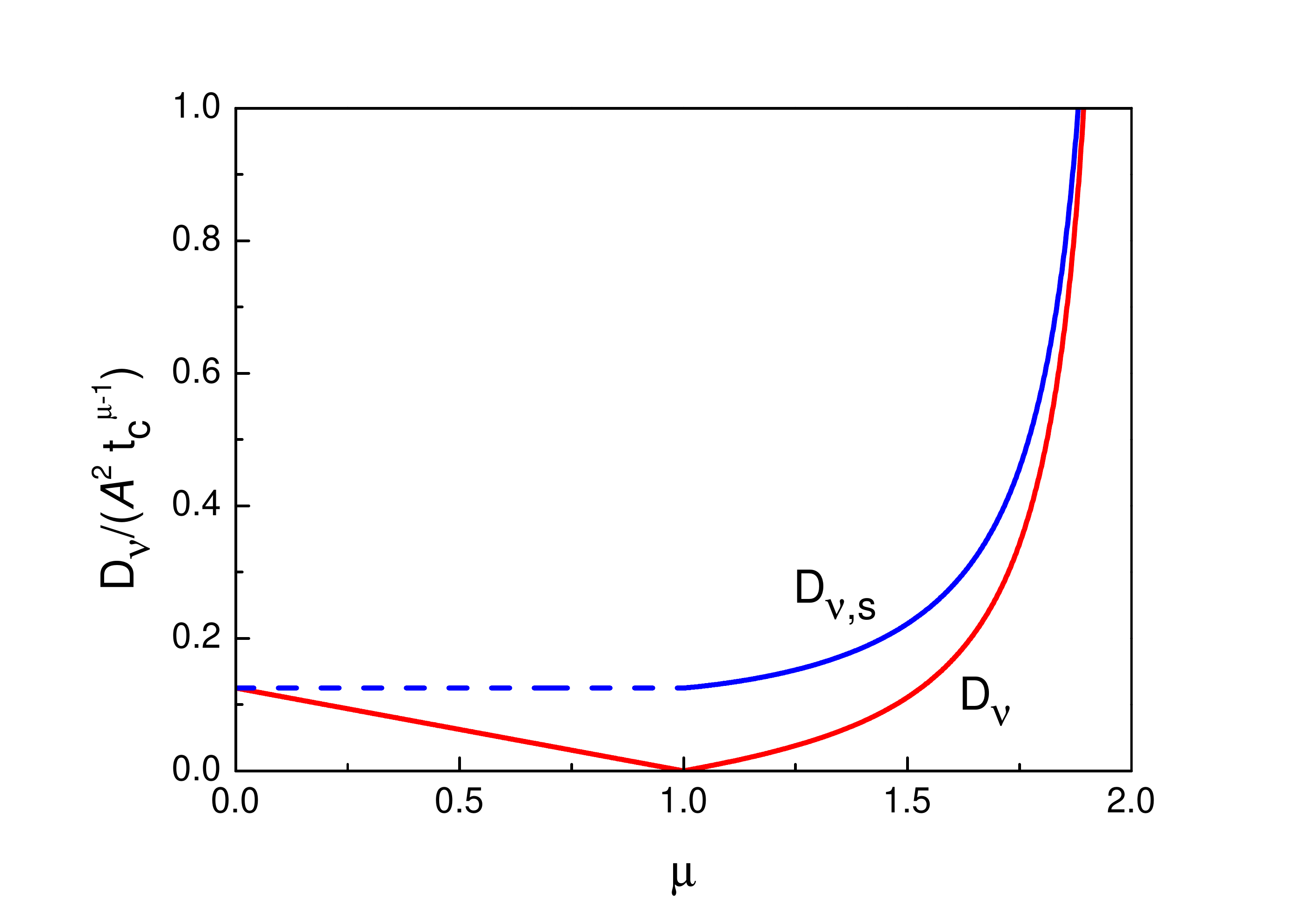}
\caption{The anomalous diffusion coefficient $D_{\nu}/(\mathcal{I}_0^2 t_{\rm c}^{\mu-1})$ for the variance of the photon number. Here we assumed a waiting time distribution of the form $\psi(t) = 0$ for $t < t_{\rm c}$ and $\psi(t) = \mu/t_{\rm c} (t/t_{\rm c})^{-\mu-1}$ for $t > t_{\rm c}$. The stationary result $D_{\nu,{\rm s}}$ (Eq.~\eqref{photvar_stat}, blue) differs from the nonstationary one $D_{\nu}$ (Eq.~\eqref{photvar}, red) but is finite at the transition to the aging phase $\mu < 1$. The blue dashed line is the result obtained for long aging times $t_0$ in the aging regime. \label{F9}}
\end{center}
\end{figure}

For $1 < \mu < 2$, we find from the general results of Ref.~\cite{mar04},
\begin{align}
\langle &I(t+\tau) I(t) \rangle \simeq \frac{1}{4} \bigg[ 1 + \frac{t_{\rm c}^{\mu}}{(\mu-1) \bar{t}} \Big[\tau^{1-\mu} - (t+\tau)^{1-\mu} \Big] \bigg], \label{levycorr}
\end{align}
which for long times $t \gg \tau$ approaches the stationary limiting form,
\begin{align}
\langle &I(t+\tau) I(t) \rangle_{\rm s} \simeq \frac{1}{4} \bigg[ 1 - \frac{t_{\rm c}^{\mu}}{(\mu-1) \bar{t}} \tau^{1-\mu} \bigg]. \label{levycorrstat}
\end{align}
Due to the constant term, these expressions are not precisely of the scaling form \eqref{6} respectively \eqref{5a}. This constant term leads to a ballistic contribution in the second moment of the photon number. However, by considering the fluctuations of the intensity $\delta I(t) = I(t) - 1/2$, which are related to the fluctuations of the photon number by $\delta n(t) = \mathcal{I}_0 \int_{0}^{t} {\rm d} t' \delta I(t)$, we can once again apply the scaling Green-Kubo relation \eqref{8} to find subballistic scaling for the variance of the photon number \cite{kla90},
\begin{align}
\langle \delta n^2(t) \rangle \simeq 2 D_{\nu} t^{\nu}
\end{align}
with $\nu = 3-\mu$ and,
\begin{align}
D_{\nu} &= \frac{\mathcal{I}_0^2 t_{\rm c}^{\mu}}{4 (3-\mu)(2-\mu) \bar{t}}, \label{photvar}
\end{align}
respectively,
\begin{align}
D_{\nu,{\rm s}} = \frac{\mathcal{I}_0^2 t_{\rm c}^{\mu}}{4 (3-\mu)(2-\mu)(\mu-1)\bar{t}}. \label{photvar_stat}
\end{align}
Similarly to our previous two examples, we have two different expressions for the anomalous diffusion coefficient depending on the initial preparation of the system.
As noted, this is in agreement with the result obtained in Ref.~\cite{zum93} using a different approach.
Both expressions for the diffusion coefficient are shown in Fig.~\ref{F9}. 
In contrast to the cases discussed in Sections \ref{SEC3} and \ref{SEC4}, the aging correlation function Eq.~\eqref{icorr} does not increase as a function of $t$ for $t \gg \tau$ but tends to a constant value. 
As a consequence the stationary diffusion coefficient does not diverge as we approach the aging regime $\mu < 1$.
For $\mu < 1$ there is no actual stationary correlation function, however, depending on whether the measurement time is large or small compared to the aging time $t_0$, we still obtain two different values for the anomalous diffusion coefficient, see Section \ref{SEC2B}.
Finally, for $\mu > 2$, the usual Green-Kubo relation is applicable and the variance of the photon number increases linearly with time.

While the explicit calculation of the intensity correlations and photon statistics is possible for the two-state model used in Refs.~\cite{jun01,mar04}, this is no longer the case for more realistic models which take into account that there may be more than a single on state or exponential cutoffs on the power law statistics of on and/or off times \cite{pet09}.
Since our scaling Green-Kubo relation Eq.~\eqref{8} only relies on the asymptotic scaling of the intensity correlation function, it can be used to relate intensity correlations and photon statistics even for models that are only tractable numerically.

\begin{figure*}
\begin{center}
\includegraphics[trim=10mm 10mm 30mm 15mm, clip, width=0.48\textwidth]{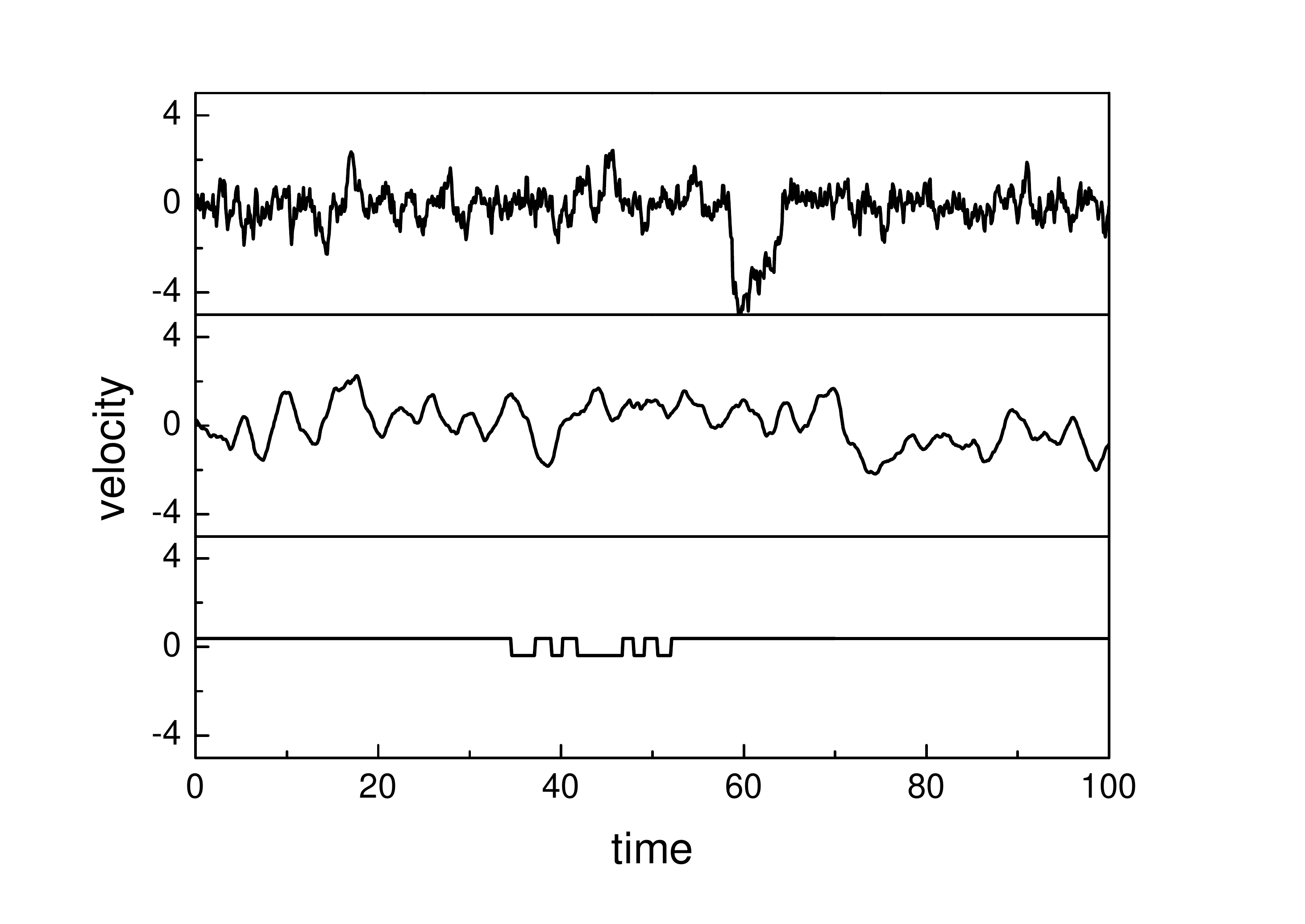}
\includegraphics[trim=10mm 10mm 30mm 15mm, clip, width=0.48\textwidth]{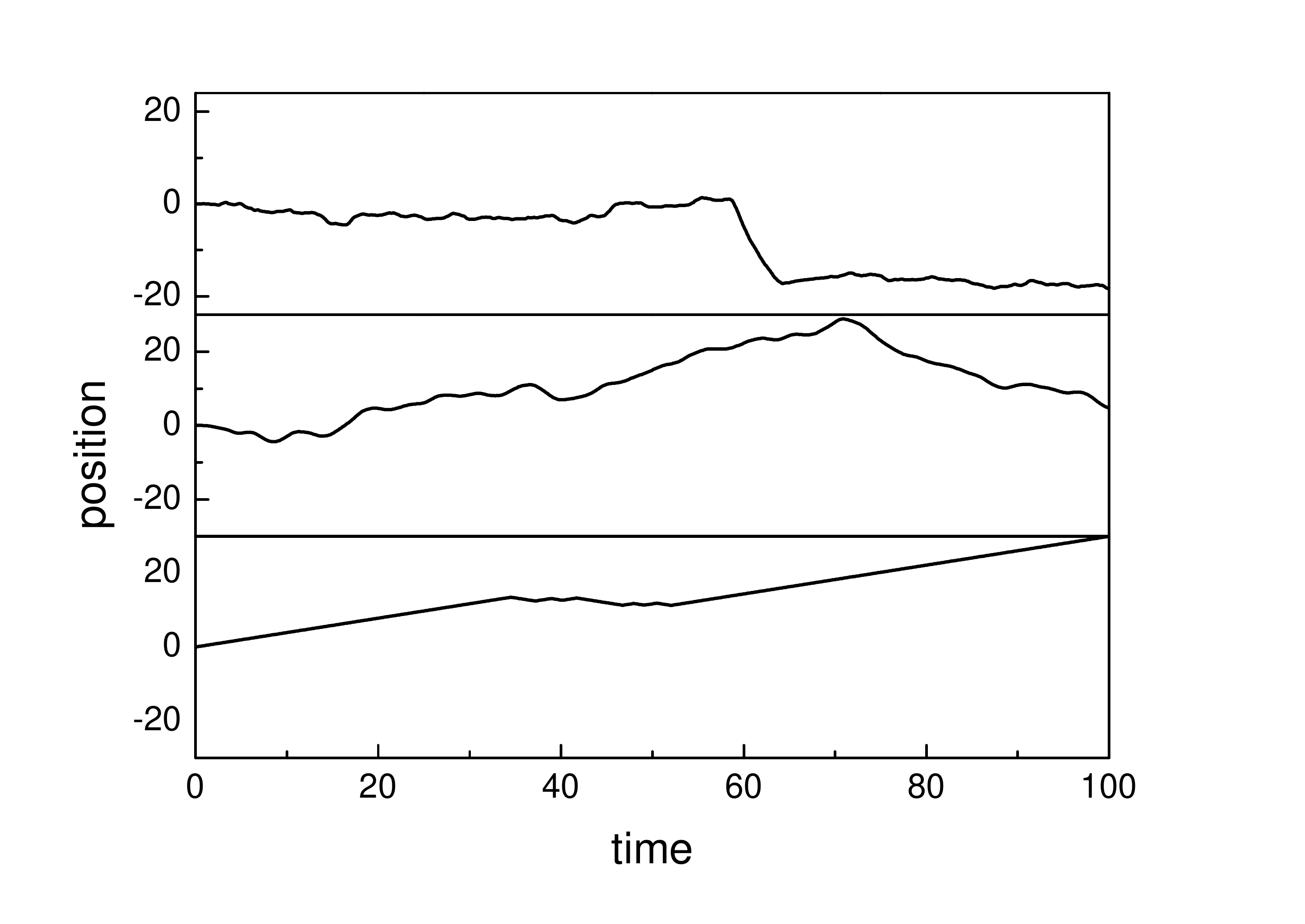}
\end{center}
\caption{Sample trajectories for the three systems discussed in Section \ref{SECIII}. Top: diffusion with a nonlinear $1/v$ friction force. Middle: fractional Langevin equation with external noise. Bottom: L{\'e}vy walk alternating between $\pm v_0$. The left figure depcits the velocity of a single trajectory as a function of time, the right figure the corresponding position. The parameters were chosen such that the diffusion exponent for all three systems is $\nu \simeq 1.5$ and the (dimensionless) anomalous diffusion coefficient is $D_{\nu} \simeq 0.25$. Despite similar diffusive properties, the actual trajectories are apparently very different. \label{F10}} 
\end{figure*}

\renewcommand{\arraystretch}{2.5}

\begin{table*}
\begin{tabular}{|c|c|c|c|c|}
\hline 
{}  & \pbox{5cm}{scaling \\ Green-Kubo} & diffusion exponent $\nu$ & correlation function & \pbox{5cm}{diffusion coefficient} \\
\hline
\pbox{5cm}{\textbf{nonlinear} $1/v$  \textbf{friction} \\ friction parameter $\frac{1}{2} < \alpha < \infty$} & $\alpha < 3$ & \pbox{5cm}{$1$ for $\alpha > 3$ \\ $4-\alpha$ for $1 < \alpha < 3$ \\ $3$ for $\alpha < 1$} & \pbox{5cm}{stationary for $\alpha > 2$ \\ superaging for $\alpha < 2$} & \pbox{5cm}{unique for $\alpha > 3$ \\ $D_{\nu}\neq D_{\nu,{\rm s}}$ for $2 < \alpha < 3$ \\ $D_{\nu} \neq D_{\nu}^{t/t_0}$ for $\alpha < 2$} \\
\hline 
\pbox{5cm}{\textbf{fractional Langevin equation} \\ friction exponent $0 < \rho < 1$ \\ noise exponent $\lambda > \rho$} & $\lambda > 2 \rho$ & $\lambda - 2\rho + 1$ & \pbox{5cm}{stationary for $\lambda < 2\rho+1$ \\ superaging for $\lambda > 2\rho+1$} & \pbox{5cm}{$D_{\nu}\neq D_{\nu,{\rm s}}$ for $\lambda < 2 \rho + 1$ \\ $D_{\nu} \neq D_{\nu}^{t/t_0}$ for $\lambda > 2\rho+1$} \\ 
\hline 
\pbox{5cm}{\textbf{Blinking quantum dots} \\ on/off time exponent $\mu > 0$} & $\mu < 2$ & \pbox{5cm}{$1$ for $\mu > 2$ \\
$3-\mu$ for $1 < \mu < 2$ \\ $2$ for $\mu < 1$} & \pbox{5cm}{stationary for $\mu > 1$ \\ aging for $\mu < 1$} & \pbox{5cm}{unique for $\mu > 2$ \\ $D_{\nu}\neq D_{\nu,{\rm s}}$ for $\alpha < 2$} \\ 
\hline 
\end{tabular}
\caption{Summary of the aging and diffusive properties for the three systems discussed in sections \ref{SEC3}, \ref{SEC4} and \ref{SEC5}. The first column lists the system and the relevant parameters. The second column states the parameter range in which the scaling Green-Kubo relation is applicable, the resulting diffusion exponent is given in the third column. The aging properties of the correlation function are given in the fourth column. The fifth column states whether the diffusion coefficient depends on the initial condition and if yes, whether it explictly depends on the aging time $t_0$. \label{tab1}}
\end{table*}

\section{Discussion \label{SEC6}}

The scaling Green-Kubo relation \eqref{8} extends the usual Green-Kubo formula \eqref{4} to the superdiffusive, nonstationary regime. 
It enables the direct evaluation of the mean square displacement $\langle x^2(t) \rangle$ from the knowledge of the scaling properties of the velocity correlation function $C_v(t+\tau,t)$. 
In particular, a scaling function replaces the stationary autocorrelation in the usual Green-Kubo formula for the determination of the diffusion coefficient. 
The scaling Green-Kubo relation is applicable to slowly decaying stationary power law correlation functions as well as nonstationary ones exhibiting aging or even superaging. 
For the latter two classes it relates, for the first time, the anomalous diffusion coefficient in a simple way to the scaling of the correlation function. 

This relation between an aging correlation function and the diffusion coefficient turns out to be important even in cases where a stationary velocity correlation function does exist but decays slowly.
Different classes of initial conditions lead to different expressions for the diffusivity: $D_{\nu,{\rm s}}$ (Eqs.~\eqref{lp_Dstat}, \eqref{FLE_Dnu_stat} and \eqref{photvar_stat}) for a stationary initial condition and $D_{\nu}$ (Eqs.~\eqref{lp_Dnu}, \eqref{FLE_Dnu} and \eqref{photvar}) if the system is not initially in the stationary state, see also Figs.~\ref{F2}, \ref{F8} and \ref{F9}.
Generally, which one of these two expressions for the diffusivity is appropriate depends on the time scales one is interested in: 
As long as the measurement time is small compared to the aging time $t_0$ (see Eq.~\eqref{GK_t0}), the anomalous diffusion coefficient is given by the stationary value $D_{\nu,{\rm s}}$. 
Once the measurement time becomes comparable to $t_0$, the diffusion coefficient will start to deviate from its stationary value, eventually reaching the nonstationary value $D_{\nu}$ for times much longer than $t_0$.
Due to this sensitivity on the initial conditions, the measurement and interpretation of anomalous transport coefficients is not as straightforward as for normal transport, where the coefficients are generally unique.
For cold atoms in optical lattices (Section \ref{SEC3}) and the fractional Langevin equation (Section \ref{SEC4}), the stationary diffusion coefficients $D_{\nu,{\rm s}}$ diverge as we approach the aging regime.
The persistence of the initial condition is directly related to the scale invariance of the velocity correlation function: There exists no typical time scale on which the correlations decay and thus the only relevant time scale is set by the aging $t_0$, which is not an intrinsic property of the system, but an external parameter determined by its initial preparation. 

Here, we discussed the persistence of the initial condition in the context of superdiffusive dynamics. Recently, the dependence of the diffusion coefficient on the initial condition has been quantified for single file diffusion, which is subdiffusive and can be related to fractional Brownian motion \cite{lei13}, indicating that this effect may be of general importance for long range correlated systems.
Another interesting question is how the sensitivity to initial conditions translates to the time-averaged diffusivity, where, instead of averaging over an ensemble of trajectories at a given time, the mean square displacement is computed from a time average over a single trajectory. The dependence of the time-averaged diffusivity on the initial conditions has so far only been discussed for special cases \cite{fro13,god13,kur13,fro13b}, and whether it will correspond to the stationary or nonstationary expression obtained here or to neither is an open question.

We have shown that the three different types of scaling correlation functions (stationary, aging and superaging) appear in a variety of physical models. 
The three examples considered in this work are of rather different types of stochastic models, yet can all be treated using the scaling Green-Kubo relation: The diffusion with a $1/v$ friction force obeys a Markovian, nonlinear, non-Gaussian Langevin dynamics, while in the case of the fractional Langevin equation, the dynamics is linear and Gaussian but non-Markovian. 
The power law blinking of quantum dots can be mapped onto a L{\'e}vy walk, which represents yet another class of stochastic model and is neither Markovian nor Gaussian.
Sample trajectories for the three models are shown in Fig.~\ref{F10}, highlighting the differences between them.
Yet due to their underlying scale invariance, the scaling Green-Kubo relation can be used in all three cases to determine the diffusive ensemble properties, which are summarized in Table \ref{tab1}. 
Our scaling Green-Kubo relation thus appears to be widely applicable and should therefore constitute an important tool for investigating the diffusive properties of many systems.

\begin{acknowledgments}
This work was supported by the Israel Science Foundation, the Focus Area Nanoscale of the FU Berlin and the DFG (Contract No 1382/4-1). AD also thanks the Elsa-Neumann graduate funding for support.
\end{acknowledgments}

\appendix

\section{Retarded and accelerated aging}

Our scaling correlation function Eq.~\eqref{6} is of the form,
\begin{align}
C(t,t+\tau) \simeq \mathcal{C} t^{\nu-2} \phi\left(\frac{\tau}{t}\right).
\end{align}
In this case, the correlation time $t_{\rm c}$ is given by the age $t$ of the system. A more general case is a correlation time that has a different functional dependence on the age, i.e. $t_{\rm c}(t)$ is some monotonously increasing function with $t_{\rm c}(0) = 0$. The corresponding correlation function is then of the form,
\begin{align}
C(t,t+\tau) \simeq \mathcal{C} t^{\nu-2} \phi\left(\frac{\tau}{t_{\rm c}(t)}\right).
\end{align}
The MSD is given by,
\begin{align}
\langle x^2(t) \rangle \simeq 2 \mathcal{C} \int_{0}^{t} {\rm d}t_2 \int_{0}^{t_2} {\rm d}t_1 \ t_1^{\nu-2} \phi\left(\frac{t_2-t_1}{t_{\rm c}(t_1)}\right).
\end{align}
We define $s = (t_2-t_1)/t_{\rm c}(t_1)$ and obtain,
\begin{align}
\langle x^2(t) \rangle \simeq 2 \mathcal{C} \int_{0}^{t} {\rm d}t_2 \int_{0}^{\infty} {\rm d} s \left(- \frac{\partial t_1(s)}{\partial s}\right) t_1^{\nu-2}(s) \phi(s), \label{x2}
\end{align}
where $t_1(s)$ is the solution of the equation,
\begin{align}
\frac{t_2-t_1}{t_{\rm c}(t_1)} = s .
\end{align}
For general $t_{\rm c}(t)$ this equation is not solvable analytically. However, the asymptotic behavior of $t_1(s)$ is given by,
\begin{align}
t_1(s) \simeq \left\lbrace \begin{array}{ll}
t_2 - s t_{\rm c}(t_2) &\text{for} \quad s \ll \frac{t_2}{t_{\rm c}(t_2)} \\
\tilde{t}_{\rm c}\left(\frac{t_2}{s}\right) &\text{for} \quad s \gg \frac{t_2}{t_{\rm c}(t_2)} ,
\end{array} \right.
\end{align}
where $\tilde{t}_{\rm c}(x)$ denotes the inverse function of $t_{\rm c}$. For the particular choice,
\begin{align}
t_{\rm c}(t) = t_{\rm a} \left(\frac{t}{t_{\rm a}}\right)^{\delta},
\end{align}
with some time scale $t_{\rm a}$ and $0 < \delta < 2$, this gives us,
\begin{align}
t_1(s) \simeq \left\lbrace \begin{array}{ll}
t_2 - s t_{\rm a} \left(\frac{t_2}{t_{\rm a}}\right)^{\delta} &\text{for} \quad s \ll \left(\frac{t_2}{t_{\rm a}}\right)^{1-\delta} \\
t_{\rm a} \left(s \frac{t_{\rm a}}{t_2}\right)^{-\frac{1}{\delta}} &\text{for} \quad s \gg \left(\frac{t_2}{t_{\rm a}}\right)^{1-\delta} .
\end{array} \right.
\end{align}
For the expression appearing in Eq.~\eqref{x2}, we have,
\begin{align}
\Big(- &\frac{\partial t_1(s)}{\partial s}\Big) t_1^{\nu-2}(s) \nonumber \\
& \simeq \left\lbrace \begin{array}{ll}
t_{\rm a}^{1-\delta} t_2^{\nu + \delta - 2}  &\text{for} \quad s \ll \left(\frac{t_2}{t_{\rm a}}\right)^{1-\delta} \\
\frac{1}{\delta} t_{\rm a}^{\frac{(\nu-1)(\delta-1)}{\delta}} t_2^{\frac{\nu-1}{\delta}} s^{\frac{1-\nu-\delta}{\delta}} &\text{for} \quad s \gg \left(\frac{t_2}{t_{\rm a}}\right)^{1-\delta} .
\end{array} \right.
\end{align}
Plugging this into Eq.~\eqref{x2}, we obtain,
\begin{align}
\langle x^2(t) \rangle &\simeq 2 \mathcal{C} \int_{0}^{t} {\rm d}t_2 \bigg[ t_{\rm a}^{1-\delta}  t_2^{\nu + \delta -2} \int_{0}^{\left(\frac{t_2}{t_{\rm a}}\right)^{1-\delta}} {\rm d}s \ \phi(s) \nonumber \\
& + \frac{1}{\delta} t_{\rm a}^{\frac{(\nu-1)(\delta-1)}{\delta}} t_2^{\frac{\nu-1}{\delta}} \int_{\left(\frac{t_2}{t_{\rm a}}\right)^{1-\delta}}^{\infty} {\rm d}s \ s^{\frac{1-\nu-\delta}{\delta}} \phi(s) \bigg].
\end{align}
The asymptotic behavior of this expression depends on the asymptotic behavior of $\phi(s)$, which we assume to be of the following form,
\begin{align}
\phi(s) \simeq \left\lbrace \begin{array}{ll}
c_{\rm l} s^{-\delta_{\rm l}} &\text{with} \quad \delta_{\rm l} < 1 \quad \text{for} \quad s \ll 1 \\[2 ex]
c_{\rm u} s^{-\delta_{\rm u}} &\text{with} \quad \delta_{\rm u} > \frac{1-\nu}{\delta} \quad \text{for} \quad s \gg 1 .
\end{array} \right.
\end{align}

{\bf The case $\delta < 1$:} For $\delta < 1$, the upper bound of the first and the lower bound of the second $s$-integral are large for $t_2 \gg t_{\rm a}$. If $\delta_{\rm u} > 1$, we may then extend the upper bound of the first integral to infinity, otherwise the integral is dominated by the value at the upper bound and thus the large-$s$ behavior of the scaling function. The $s$-integral in the second term only depends on the large-$s$ behavior. We then have,
\begin{align}
\langle &x^2(t) \rangle \simeq 2 \mathcal{C} \nonumber \\
& \times \left\lbrace\begin{array}{ll}
\frac{1}{\nu+\delta-1} t_{\rm a}^{1-\delta} t^{\nu+\delta-1} \int_{0}^{\infty} {\rm d}s \ \phi(s) &\text{for} \quad \delta_{\rm u} > 1 \\[2 ex]
\frac{c_{\rm u} (1+\frac{1}{\delta})}{\nu + (\delta-1)\delta_{\rm u}} t_{\rm a}^{(1-\delta)\delta_{\rm u}} t^{\nu+(\delta-1)\delta_{\rm u}} &\text{for} \quad \delta_{\rm u} < 1.
\end{array} \right.
\end{align}
We thus have two different results for the diffusion coefficient and exponent depending on the large-$s$ behavior of the scaling function: For $\delta_{\rm u} < 1$, i.e. when the scaling function $\phi(s)$ decays slowly or even increases at large $s$, the diffusion coefficient is determined by this large-$s$ behavior. This agrees with the physical intuition, since for $\delta < 1$, the system ages more slowly and thus the low-age behavior (corresponding to $t \ll \tau$ respectively large $s$ in the correlation function) is important. For $0 < \delta_{\rm u} < 1$ the diffusion is retarded (i.e. the diffusion exponent is smaller than $\nu$) while it is accelerated for $\delta_{\rm u} < 0$. On the other hand, if the scaling function decays very fast ($\delta_{\rm u} > 1$), the contribution from the low-age regime is numerically small and of the same order as the high-age one. Thus, the diffusion coefficient depends on detailed shape of $\phi(s)$ and not only on the large-$s$ expansion. In this case the diffusion is always retarded and the system may even become subdiffusive for $\delta < 2-\nu$.

{\bf The case $\delta > 1$:} For $\delta > 1$, the upper bound of the first and the lower bound of the second $s$-integral are small for $t_2 \gg t_{\rm a}$. If $\delta_{\rm l} < (1-\nu)/\delta$, we may then extend the lower bound of the second integral to zero, otherwise the integral is dominated by the value at the lower bound and thus the small-$s$ behavior of the scaling function. The $s$-integral in the first term only depends on the small-$s$ behavior. We then have,
\begin{align}
\langle &x^2(t) \rangle \simeq 2 \mathcal{C} \nonumber \\
& \times \left\lbrace\begin{array}{ll}
\frac{\delta }{\nu+\delta-1} t_{\rm a}^{\frac{(\nu-1)(\delta-1)}{\delta}} t^{\frac{\nu+\delta-1}{\delta}} \int_{0}^{\infty} {\rm d}s \ s^{\frac{1-\nu-\delta}{\delta}} \phi(s) \\
\qquad \qquad \qquad \text{for} \quad \delta_{\rm l} < \frac{1-\nu}{\delta} \\
\frac{c_{\rm u} (1+\frac{1}{\delta}) }{\nu + (\delta-1)\delta_{\rm l}} t_{\rm a}^{(1-\delta)\delta_{\rm l}} t^{\nu+(\delta-1)\delta_{\rm l}} \\
\qquad \qquad \qquad \text{for} \quad \delta_{\rm l} > \frac{1-\nu}{\delta}.
\end{array} \right.
\end{align}
Similarly to the case $\delta < 1$, we find two qualitatively different behaviors, though now depending on the small-$s$ expansion of the scaling function, as the system now ages faster and thus the high-age behavior is important. For $\delta_{\rm l} > (1-\nu)/\delta$, the scaling function diverges at small $s$ or slowly tends to zero and the high-age behavior ($t \gg \tau$ or small $s$) dominates the diffusion coefficient. Depending on the sign of $\gamma_{\rm l}$ the diffusion may be retarded or accelerated. For $\delta_{\rm l} < (1-\nu)/\delta$, the scaling function rapidly tends to zero for small $s$ and thus the high-age contribution and the low-age one are of the same order. As before, the diffusion coefficient depends on the entire domain of $\phi(s)$.

For $\delta = 1$, both the low-age and high-age contributions are always of the same order and the behavior of $\phi(s)$ over the whole range of its argument determines the diffusion coefficient. This underlines the special importance of the "true" scaling type of correlation functions with $\delta = 1$ discussed in the main body of the paper and in the examples.

\FloatBarrier

\bibliography{GGK}

\end{document}